\documentclass[10pt]{JHEP3}
\pdfoutput=1 

\usepackage{amsmath}
\usepackage{verbatim}
\usepackage{graphicx}
\usepackage{cleveref}

\newcommand{\q}{q^{\ast}}

\newcommand{\be}[1]{ \begin{equation}\label{#1} }
\newcommand{\ee}{\end{equation}}
\newcommand{\bea}{\begin{eqnarray}}
\newcommand{\eea}{\end{eqnarray}}

\newcommand{\p}{\partial}

\newcommand{\mbz}{\mathbb{Z}}
\renewcommand{\t}{\tau}
\newcommand{\non}{\nonumber}

\renewcommand{\l}{\left(}
\renewcommand{\r}{\right)}

\title{Higher Spin de Sitter Quantum Gravity}
\author{Rudranil Basu $^{a,b}$
\\$^a$Physique Th\'{e}orique et Math\'{e}matique
\\ Universit\'{e} Libre de Bruxelles and International Solvay Institutes.
\\ Campus Plaine C.P. 231, B-1050 Bruxelles, Belgium  \\ $^b$ Indian Institute of Science Education and Research, 
\\ Dr. Homi Bhabha Road, Pashan, Pune 411008. India.
\\{}\\ Email: \email{rudrobose@gmail.com}}
\preprint{}
\abstract{We consider Einstein gravity with positive cosmological constant coupled with higher spin interactions and calculate Euclidean path integral perturbatively. We confine ourselves to the static patch of the 3 dimensional de Sitter space. This geometry, when Euclideanlized is equivalent to 3-sphere. However, infinite number of topological quotients of this space by discrete subgroups of the isometry group are valid Euclidean saddles as well. The case of pure Einstein gravity is known to give a diverging answer, when all saddles are included as contribution to the thermal partition functions (also interpreted as the Hartle Hawking state in the cosmological scenario). We show how higher spins, described by metric-Fronsdal fields help making the partition function finite. Counter-intuitively, this convergence is not achieved by mere inclusion of spin-3, but requires spin-4 interactions.}
\begin{document}

\baselineskip 3.5ex

\section{Introduction}
The main question we would be seeking answer for, in the following analysis is the behaviour of quantum gravity partition function on de Sitter (3 dimensional) space and (almost all of) its factors by an infinite class of discrete groups. The motivation behind this is primarily to look for correlation functions for quantum field theories on de Sitter space, when minimally coupled quantum gravity fluctuations are included. 
However that would require an understanding of a vacuum state. While it is understood that isometries of (global) de Sitter space would fix the vacuum state for the field theory, the notion of such a state is not very clear for quantum gravity fluctuations. Any theory of gravity should possess time reparametrization invariance resulting the Hamiltonian to vanish as a constraint. This would prevent us from identifying the vacuum state as a minimal energy configuration. But it was the seminal work by Hartle and Hawking \cite{Hartle:1983ai} where a plausible candidate for the same was proposed by defining the vacuum state as a Euclidean functional integral over geometries with fixed data over some compact co-dimension-1 hyper-surface.

To facilitate computation in the functional integral approach, we resort to Euclidean formulation. As a background, Eulclideanized version of the static patch of it will be considered. This space can be identified isometrically with $S^3$. Now, while investigating the Hartle-Hawking vacuum state for quantum gravity in the functional integral approach, it is natural to include all classical saddles (as fixed data) or geometries, around which fluctuations would be evaluated in a perturbative manner. An extremely well-understood fact is that, 3D Einstein gravity without matter does not have propagating degrees of freedom in the bulk. Any local description of it with finitely many degrees of freedom at each point of space-time comes along with exactly same finite number of constraints coming from gauge redundancy. It implies that a large number (infinite) of gauge inequivalent class of degrees of freedom, if possible, can only come from boundary, as in the case of AdS${}_3$. But for closed and compact background manifolds like $S^3$ or its quotients, only finite number of topological degrees of freedom can be possible, if at all. For example, we see that the only geometry, that is smooth on $S^3$ is:
\bea \label{s3}
 d\,s^2 = d\,r^2 + \cos ^2 r \,d\t ^2 + \sin^2 r \,d\ \phi^2
\eea
Range of $r$ is fixed ($\in [0, \pi/2]$) such that it covers the Euclideanization of the static patch of dS${}_3$, not the global de Sitter. Moreover it is clear that smoothness of the above geometry near $r=0$ and $r= \pi/2$ demand the following identification:
\bea \label{idens3}
 (\tau, \phi ) \equiv (\t, \phi )+ 2\pi (m,n) \quad \mbox{ for } \quad m, n \in \mathbb{Z} . 
\eea

If we wish to include more saddles, we would have to look for manifolds which allow locally smooth de Sitter (or $S^3$) geometries, but are topologically distinct from $S^3$. These are quotients like $S^3/ \Gamma$ where $ \Gamma$ is some finite group and $ \Gamma \subset SO(4)$. Here $SO(4)$ is the isometry group of $S^3$. These spaces are topologically different for different choices of $ \Gamma$. The homotopy groups of these spaces have the same order as that of $\Gamma$. For the action of $\Gamma$ on $S^3$ to be free and discrete, it must be one of the polyhedra groups. Abelian candidates for $\Gamma$ are obviously the cyclic groups $ \mathbb{Z}_p , \, p \in \mathbb{N}$. Although there are other groups, like the dihedral group and other crystallographic subgroups of $SO(4)$, for the present case, we would be content with $\mathbb{Z}_p$ for reasons which would be evident as we progress. Let us now focus on the interesting mechanism of how the local geometries of the topological quotients of $S^3$ get induced from that of $S^3$. 

This can best be understood by embedding $S^3$ in $\mathbb{C}^2$ by:
$$ |z_1|^2 + |z_2|^2 =1 .$$
Action of $\Gamma = \mathbb{Z}_p$ on $S^3$ is free and can then be defined as:
$$\Gamma: (z_1, z_2) \rightarrow (e ^{\frac{2 \pi  i}{p}}\, z_1, e ^{\frac{2 \pi \,q\, i}{p}}\, z_2),$$ for a positive integer $q < p$ which is coprime to $p$. We will use the notation $(q,p)=1$ for denoting that $q$ and $p$ don't have any non-trivial common factor. Now let us choose the coordinates of \eqref{s3} such that $z_1 = \cos r \,e^{i \t}, z_2 =\sin r \, e^{i \phi}$. On $S^3/ \Gamma$ it would mean the identification :
\bea \label{idenlens}
 (\t, \phi) \sim (\t, \phi) + 2 \pi\,(\frac{m}{p}, \frac{m\,q}{p} + n) \quad \mbox{ for } \quad m,n \in \mathbb{Z} \eea 
most generically. The resulting quotient, now depending upon both $q, p$ is named a Lens space, $L(p,q)$. It is easy to see, as $\Gamma$ acts freely on $S^3$ and that the parent space $S^3$ was a smooth embedding in $\mathbb{C}^2$, the induced geometry on $L(p,q)$ is also smooth. 

\subsection{More on the topological structure of Lens spaces}
We just discussed Lens spaces as quotients of $S^3$ and how the geometry of $S^3$ (or thermal de Sitter) induces the geometry on them. While doing so, we keep in mind that the induced geometry (metric) is a local structure, while global properties are captured in the topological properties. 

In this view, let us stress here that a Lens space can also be constructed by gluing two solid tori at their boundaries (which are 2-tori themselves). While this gluing prescription is not unique, it is neither completely arbitrary. One uses an element of the mapping class group $PSL(2, Z) \equiv SL(2, \mathbb{Z}) /\mathbb{Z}_2$ (also in this case the group of modular transformations) of the toric boundaries. Another structure of relevance here would be the fundamental group $ \pi _1(T^2) = \mbz \oplus \mbz$ of the boundary tori. Hence any cycle on the boundary tori can be represented as words of the form $a^m \circ b^n; m,n \in \mbz $, where $a$ and $b$ are the fundamental cycles \footnote{The fundamental group of the toric boundary, in absence of any punctures, has an additional feature of being abelian; \textit{ie} $a\,b\,a^{-1}\,b^{-1} =1$}. One of these, say $a$ for example, becomes contractable when the bulk of the solid torus is taken in to account. 

To illustrate the gluing mechanism, let us start with two solid tori and consider the non-contractible $b$ cycles on each of them. If we use 
the action of the element
\[ S=\left( \begin{array}{cc}
0 & -1  \\
1 & 0  \end{array} \right) .\] on the $b$ cycle of one of the solid tori, it would transform it to $a$, now a contractable one, $S: b \rightarrow a$. Now gluing these solid tori at their boundaries with identifying $b$ of one of them to the transformed one on the other would produce $S^3$, the most trivial Lens space.

In order to generate more non-trivial Lens-spaces, other elements of the modular group come of use. More precisely, when the element:
\[ U_{p,q}=\left( \begin{array}{cc}
q & b  \\
p & d  \end{array} \right) \in PSL(2, \mbz)\]
is used, the Lens space $L(p,q)$ is generated. As is clearly seen that $U_{p, q}$ is not unique, given only the values of $p$ and $q$. Hence the map from $PSL(2, \mbz)$ to set of all Lens spaces is not one-to-one but is non-injective surjection. Rather the left right coset $ \mbz \backslash PSL(2, \mbz) / \mbz$ is the subgroup which has injective maps to the set of Lens spaces. 


\subsection{Thermal properties and comparison with thermal AdS}
Let us remind ourselves that Euclideanization would mean compactifying the time direction to an angular coordinate. This converts 3D de Sitter to $S^3$. In effect, field theory correlators would be equivalent to ones evaluated in a canonical ensemble at temperature ($ \beta =2 \pi$) defined by the periodicity of the Euclidean time. On the other hand, the correct ensemble for the $S^3/\mathbb{Z}_p \equiv L(p,q)$ which is defined by the identification \eqref{idenlens} would be a grand canonical one with inverse temperature $ \beta = \frac{2 \pi }{p}$ and angular potential $ \mu = \frac{2 \pi \,q}{p}$. This can be described by the thermal state 
$$ \rho = e^{- \beta \,H + i \mu \, J} $$ where $H$ and $J$ generate Euclidean-time translation and rotation along the vector fields $ \p_{\tau}$ and $ \p_{\phi} $ respectively. It's interesting to note that both the temperature and the angular potential take discrete values, each of which define a Lens space. As $p$ grows large a continuum of sort is however reached, which are of decreasing volume \footnote{Volume of a space $S^3/\Gamma$, with respect to some metric on $S^3$ is $\mathrm{Vol}(S^3)/|\Gamma|$. The order $|\Gamma|$ for the discrete group $\mathbb{Z}_p$ is of course $p$.} as well. This very curious thermal property exhibited by thermal dS is only exhibited by the saddles $ S^3/\mathbb{Z}_p$. The other saddles found by factoring $S^3$ by other possible discrete subgroups of $SO(4)$ don't have this beautiful thermal interpretation \cite{Castro:2011ke}. Therefore the physical meaning of the Euclidean partition function is not very clear in that case. This is principally the reason for not taking those saddles into consideration while calculating 1-loop partition function for quantum gravity or other excitations on thermal de Sitter background.

We also point out here that The choice of Hartle Hawking as a vacuum is practically useful, particularly from Euclidean gravity point of view. However one must keep in mind that this description of a field theory vacuum requires the equilibrium cosmological horizon to be kept at a constant temperature $ \beta = 2 \pi$. Pair production of particles must be described by the field theory in order to maintain the horizon at equilibrium. On the other hand their is another choice of a vacuum state, namely the Bunch Davies vacuum, which can be defined through data in far past, hidden inside the cosmological horizon.

This is quite in contrast to the well-understood AdS case. Three dimensional Euclidean AdS space in identified topologically with the bulk of a solid torus. The appropriate thermodynamic ensemble that describes the system, is given by the modular parameter $ \tau$ of the boundary torus. $ \left(\mathfrak{Im}(\t)\right)^{-1}$ gives the temperature whereas $ \mathfrak{Re}(\tau)$ the angular potential of the system. Now, $\t$ varies continuously in the fundamental domain of the upper half plane. Hence, the thermodynamic parameters take continuous values, as expected. 

The partition function calculated for the system with a given $\t$  would correspond to a thermodynamic one. But this must be modular invariant to be physically meaningful. In order to achieve modular invariance of the partition function and hence all derived quantities from it, one takes sum over all modular images of the thermal AdS. This is done by acting the thermal AdS (Eucledian) modular parameter with the modular group $PSL(2, \mathbb{Z})$ (a subset of it, to be precise), much in the sense of the Farey tail expansion \cite{Dijkgraaf:2000fq}. For example, the saddle, which corresponds to thermal AdS is related to the one corresponding to Euclidean BTZ solution through an $S :  \t \rightarrow -\frac{1}{\t}$ modular transformation.

On route defining the Hartle Hawking state we will be summing over all Lens spaces in our case. In spirit of the discussion made in the last subsection, we see that this effectively means summing the partition function over a subset of the modular group, which has injective maps to the space of Homeomorphically distinct Lens spaces.
\subsection{Hartle-Hawking state as a sum over Lens spaces with higher spins}

As mentioned earlier, our approach for studying the Hartle-Hawking vacuum state would be through the functional integral evaluation. Alternatively it boils down to calculate the thermal partition function in an appropriate ensemble. We would be interested in a perturbative analysis over classical saddles. The infinite family of quotients of thermal dS, ie the Lens spaces serve as the classical saddles, all sharing the same local gravitational configuration and differing only in topology.

In section 2, we describe how the partition function diverges as one takes into account the whole family of Lens spaces, following the analysis of \cite{Castro:2011xb}. This divergence essentially indicates instability of the vacuum state. However we must mention that in a subsequent paper \cite{Castro:2011ke} by the same authors a prescription for finite partition function and hence physical vacuum was given. In this later case they considered propagating massive graviton modes on top of pure topological gravity. An alternative route for showing finiteness of the partition function using methods of topological QFT was pursued in \cite{Basu:2011vs}. In spite of these analyses, the origin of the divergence was not very clear from a physical point of view, neither was the point that more degrees of freedom may wash out the divergence. Nature of the divergence is through the harmonic series in $p$ as we some over all Lens spaces $L(p,q)$. Effectively this is logarithmic in the discrete variable $p$. Note that, with respect to the de Sitter metric \eqref{s3}, volume of $L(p,q)$ varies as $p^{-1}$.

Now looking at the nature of the divergence which is extremely slow in volumes of the Lens spaces, it may be tempting to say that any small perturbation in the effective (quantum corrected) volume would cure the divergence. However, that still is not the case, as there is nothing which can be ascribed as a \textit{slowest diverging series}. One safe way to get rid of this divergence would be through some higher (than 1) inverse power of $p$ contribution of functional integral on $ L(p,q)$. We then take clue from the structure of the partition function in the AdS case \cite{Gaberdiel:2010ar, Gaberdiel:2010pz} where one sees the linearized higher spin fluctuations of similar form contribute to the 1-loop partition function multiplicatively. As we will see in section 3 in the present paper, that inclusion of higher spins indeed resolves the divergence through a faster decay of the functional integrals in inverse powers of $p$. Higher spins in 3 space-time dimensions has its virtues worth studying. One can couple finite number of higher spins with pure spin-2 gravitational interaction consistently, quite in contrast to 4 dimensional higher spin Vassiliev theory. Secondly, higher spin interactions do not bring along with it any local degrees of freedom. 
In effect this retains the topological nature of the theory. Finally the famous no-go theorems prohibiting higher-spin interactions in flat space (vanishing cosmological constant) do not apply in 3 dimensions, allowing one to explore these in flat 3 dimensional space as well \cite{Afshar:2013vka}.

Study of the wave-function of the 3d quantum universe, ie Hartle-Hawking state on de Sitter space in other avenues, especially through dS/CFT correspondence \cite{Strominger:2001pn, Strominger:2001gp}, been made to some extent in recent past \cite{Castro:2012gc}. These studies essentially require a subtle \cite{Maldacena:2002vr} analytic continuation of de Sitter space so as to map to 3 dimensional AdS space. The advantage in this approach is that there exist a large number of results known in  in the AdS context, which can be mapped back to the case of de Sitter. For example the holographic renormalization technique, discovered in the AdS/CFT context has been successfully applied to 3 dimensional dS/CFT correspondence using this analytic continuation technique, in presence of higher spin fields as well \cite{Lal:2012py}. Interestingly, the Hartle-Hawking wave-function derived \cite{Castro:2012gc} starting from AdS partition function comes out to be non-normalizable, indicating a non-perturbative de Sitter instability. Although, outside the scope of the present paper, it would be interesting to see what effect higher spin fluctuation has to say on this on this instability. It is worthwhile to mention here, that the same approach of mapping with AdS has resulted into resolving cosmological singularities \cite{Krishnan:2013cra} and uncovering of interesting de Sitter backgrounds \cite{Krishnan:2013zya} allowing higher spins at the classical level. Again this is not of immediate concern to our work, since we confine to the boundary-less compactification of static patch of de Sitter only.

On the other hand similar question, not exactly the same though, has been pursued in detail in 4 space-time dimension, in the context of developing dS/CFT correspondence \cite{Anninos:2011ui, Anninos:2012ft, Anninos:2013rza}. A concrete 3d CFT example has been constructed with $Sp(N)$ anti-commuting scalars dual to Vasiliev higher spin gravitational degrees of freedom in the bulk. This stands as the de Sitter analogue of the $O(N)$ model CFT${}_3$ by Giombi-Yin \cite{Giombi:2011ya}. In particular, in \cite{Anninos:2013rza}, the wave-function constructed from the particular CFT dual showed a feature of instability for zero mode of the mass `function' of the boundary $Sp(N)$ scalars. Interestingly, attempts at looking for CFT duals corresponding to the static patch of four dimensional de Sitter space have also been made, particularly in presence of higher spin fluctuations \cite{Karch:2013oqa}.

In this light it would have been ideal to go on probing for a CFT  in the lines of Gaberdiel-Gopakumar duality \cite{Gaberdiel:2010pz} at the boundary of 3d de Sitter, which would stand as yet another example of dS/CFT. However the compactification that we perform here for the Euclideanization of static patch of de Sitter or its quotients and that suits our primary aim of computing higher spin functional determinant, does leave us with boundary-less manifolds. We defer to last section of this paper for a more elaborate discussion of this possibility and expectations from such an analysis. We should rather stress the point that the reason for the success story of higher spin fluctuations in the present case is not very clear physically. The dual CFT spectrum, if uncovered would prove to be useful in this question.

\subsection{Chern Simons picture} 
The gauge theoretic formulation of 3D gravity in terms of Chern Simons theory \cite{Witten:1988hc} is well-understood now, even including the fact where the correspondence of gravity described by metric formulation and first order connection formulation fails \cite{Witten:2007kt}. One beautiful aspect of the work \cite{Castro:2011xb} is that, this shows how the 1-loop partition function in the metric formulation matches with the one found in Chern Simons approach. The later computation requires borrowing results from topological quantum field theory literature \cite{Jeffrey:1992tk}, particularly concerning Lens space topology. However the CS partition function does not naively match with the metric answer, as there are a number of saddles in the gauge theory side for a given topology $L(p,q)$. Not all of these saddles are physically meaningful, in the sense that not all of them do correspond to smooth Riemann metric. This prevents them from being locally isomorphic to Euclidean de Sitter.

There is one advantage in working in the gauge theory picture. That is the full non-perturbative answer is known for Chern Simons partition function on Lens spaces. CS theory, which is devoid of propagating degrees of freedom, is known to be 1-loop exact in many cases. However on Lens spaces, it is two loop exact and with instantons. Moreover, it is not very well understood how the gravity theory should be described in terms of this non-perturbative answer. Neither is it clear if it should still be equivalent to the Chern Simons one at the non-perturbative level. Another important aspect of the CS quantization is that it background independent, ie does not require a background metric on which quantum fluctuations may take place. This is in contrast with the perturbative metric theory route.

The gauge group for CS in presence of positive cosmological constant is $SU(2) \times SU(2)$, in Euclidean \footnote{The phrase Euclidean signature may be a bit confusing, as we have already mentioned that CS is independent of background metric. By Euclidean, we mean here that the triad dreibein that describes the first order gravity theory, transform in the adjoint representation of $SU(2)$. On the other hand Lorentzian would mean it transforming under $SO(2,1)$.} signature. It would be natural to consider the triad and connection-like fields for higher spin fields to be arranged such that the Chern Simons gauge group denoting spin-$N$ fields to be $SU(N) \times SU(N)$. Using the methods developed in \cite{Hansen:2003ky}, one can compute non-perturbative partition functions for those theories on Lens spaces. The Chern Simons result reconfirms that higher spin interactions cure the divergence problem. However extracting out the semi-classical result from the non-perturbative one and choosing the appropriate saddle describing the physical geometry is an extremely interesting crux, which is open for future work. It is still not very well understood as to how the conditions of classically allowed saddles can be implementedon the non-perturbative partition function of CS, particularly in presence of higher spins. As this matter is going to be a part of a separate investigation on its own, we have not included this analysis in the main body of the work.

\subsection{Plan of the paper}
First we review the the case of pure spin-2 gravity and analyze the issue of divergence in the Hartle-Hawking state or the partition function and hint at possibilities how this may be removed by inclusion of higher spin fluctuations. In section 3 we consider linearized spin-3 theory to systematically find the determinant that has to be evaluated for 1-loop partition function. Using known results about the spectra of the Elliptic operators for a given Lens space, we go on evaluating the determinants and employ standard regularization techniques to get meaningful answer. Moreover we calculate the contribution of zero modes in section 3.3. Then explain the need of spins higher than 3 for a converging sum over all Lens spaces. The case of spins greater than equal to 4 is considered in section 4 to reach the desired convergence issue. We have drawn a somewhat brief but self-consistent picture of the Chern Simons side of the analysis in a section in Appendix (app C).
\section{The ailment of spin 2 pure gravity}
In \cite{Castro:2011xb}, the perturbative partition function was calculated up to 1-loop quantum fluctuations for pure gravity in de Sitter background both in metric variables and in the first order vielbein-connection formalism. Let us for the moment, review their result in the regime of metric formalism solely. The Euclidean action is given as:
\bea \label{sp2}
S[g] = -\frac{k}{4 \pi}\int \, \sqrt{g}\, (R -2) \, d^3 x
\eea
with $k = \frac{1}{4G}$. We have chosen to set the cosmological constant to be 1 here.
As expected from our previous discussion, the full partition function was represented as
\begin{eqnarray}
 \label{def}
Z = \sum_{q,p} Z^{[\mathrm{tree}]}_{(2)}(p,q) \, Z^{[\mathrm{1-loop}]}_{(2)}(p,q)
\end{eqnarray}
where $Z^{[\mathrm{tree}]}_{(2)} (p,q)$ and $Z^{[\mathrm{1-loop}]}_{(2)} (p,q)$ respectively are the tree level and the one-loop partition function on $L(p,q)$. The subscript $2$ denoting spin-2 fluctuations although is redundant here, would be important when more spins will be incorporated. The sum is over all Lens spaces. Background geometry is chosen as \eqref{s3} with the identification \eqref{idenlens}.

Let us first scrutinize what the tree level contribution of all saddles tell us (see \cite{Castro:2011xb} section (3.2)). In this view we momentarily ignore the effect of first quantum corrections and use:
\begin{eqnarray}
 \label{deftree}
Z &=& \sum_{q,p}  Z^{[\mathrm{tree}]}_{(2)} (p,q)\non \\
  &=& \sum_{q,p} \exp \left(- S[g]\biggr\rvert_{L(p,q)}\right) = \sum_{q,p} \exp \left( \frac{2 \pi k}{p}\right)\non \\ 
  &=& \sum_{p=1}^{\infty} \phi(p) \, \exp \left( \frac{2 \pi k}{p}\right),
\end{eqnarray}
where $\phi(p) $ is the Euler totient function  
\bea \label{totient}
\phi(p)=\sum _{ \substack{q( \mathrm{mod}\, p)\\ (q,p)=1}}1 .\eea That \eqref{deftree} diverges, can be seen easily:
\bea \label{zet}
\sum_{p=1}^{\infty} e^{2 \pi k /p} \phi(p) &=& \sum_{p=1}^{\infty} \sum_{s=1}^{\infty} \frac{(2 \pi k)^s}{s!} \, \frac{\phi(p)}{p^s} \non \\
&=& \sum_{s=1}^{\infty} \frac{(2 \pi k)^s}{s!} \, \frac{\zeta(s-1)}{\zeta(s)}.
\eea 
Here the `zeta function' for totient function $\sum_{p=1}^{\infty} \dfrac{\phi(p)}{p^s}$ was expressed in terms of ratio of Riemann zeta functions. We observe that the term for $s=2$ in the above sum deiverges, as this corresponds to singularity of $\zeta(1)$ in the numerator. Moreover as all the terms are positive, there is no chance of phase-cancellation and making this sum finite.
\footnote{ Interestingly, in their un-regularized form,
$$\sum_{p=1}^{M} \phi(p) \sim \frac{3 M^2}{\pi^2}, ~~ \sum_{p=1}^{M} \frac{\phi(p)}{p} \sim \frac{6 M}{\pi^2}$$
as $M$ grows large.}

The situation does not improve either, even after including 1-loop quantum corrections. For a particular Lens space $L(p,q)$, the partition function up to 1-loop (for $q \neq p-1$) is calculated \footnote{Note that the corresponding equation, \textbf{(4.31)} of \cite{Castro:2011xb} has a typing error in form of the factors of 2 in the sine functions.} to be:
\bea \label{ls1}
\frac{8 \pi}{k p} \, e^{2 \pi k/p} \sin\left( \pi\dfrac{q+1}{p}\right) \sin\left( \pi\dfrac{q-1}{p}\right)\sin\left( \pi\dfrac{q^{ \ast } + 1}{p}\right)\sin\left( \pi\dfrac{q^{ \ast } - 1}{p}\right)
\eea
where $q^{\ast}$ is a unique positive integer modulo $<p$ with the property: $q\, q^{\ast} =1 \, \mbox{mod}\, p$. We have stated the above 1-loop result without giving a derivation, which can be found in \cite{Castro:2011xb}. However, this would be included in our higher spin analysis automatically, which we aim to exhibit next. Also note that the $1/k$ factor introduced through contributions of zero modes \cite{Carlip:1992wg} and is a non-perturbative effect. A quick comparison of the 1-loop result \eqref{ls1} with the tree-level one reveals that quantum effects indeed slows the divergence. However, divergence cannot be avoided completely as to be seen below.

Using trivial manipulations on \eqref{ls1} we encounter the worst (in terms of behaviour at large $p$) term:
$$ \dfrac{e^{2 \pi k/p}}{p} \cos^2 \left(\frac{2 \pi}{p}\right).$$
When this term is summed over all Lens spaces $L(p,q)$ with fixed $p$, one sees the appearance of the totient function again as in \eqref{zet}:
\bea \label{cosdiv}&&\sum _{p =1} ^{ \infty} \dfrac{e^{2 \pi k/p}}{p} \cos^2 \left(\frac{2 \pi}{p}\right) \, \phi(p) \non \\
&=& \frac{1}{2} \sum _{m=0}^{ \infty} \frac{\l 2 \pi k\r ^m}{m!} \sum_{p=1}^{\infty} \phi(p) \l p^{-m-1} +  \sum_{n=0}^{\infty} \l -1 \r ^n \frac{\l 4 \pi \r ^{2n}}{(2n)!} p^{-m-2n-1} \r \non \\
& = & \frac{1}{2} \sum _{m=0}^{ \infty} \frac{\l 2 \pi k\r ^m}{m!} \l \frac{\zeta (m)}{\zeta(m+1)} + \sum_{n=0}^{\infty} \l -1 \r ^n \frac{\l 4 \pi \r ^{2n}}{(2n)!} \frac{\zeta(m+2n)}{\zeta(m+2n+1)} \r.\eea
A close inspection of the last line reveals that the term $m=1, n=0$ gives rise to $ \dfrac{\zeta(1)}{\zeta(2)}$, which diverges. 
This series diverges as per the analysis carried out it in \cite{Castro:2011xb}. The same $ \zeta(1)$ term as in the tree level analysis appears here as the reason of divergence.

In order to look for a resolution from this problem, we notice that a faster fall-off of the series (first line of \eqref{cosdiv}) than $p^{-1}$ has a chance of saving the sum. In case we have a theory which would result in a 1-loop partition function with a dependence of $p^{-2}$, that would still not be good enough to make it convergent. Because even in that case the diverging $ \zeta(1)$ would still make its appearance through the term $m=0=n$ in a computation analogous to \eqref{cosdiv}. Therefore a 1-loop partition function for a Lens space $L(p,q)$ of the form
\bea \label{formpart}
Z \sim \frac{e^{2 \pi k/p}}{p^l}
\times \mbox{ Trigonometric functions}(q, q^{\ast},p)
\eea
for $l \ge 3$ would clearly be able to steer clear of the diverging zeta function.

But we now face the question: what would make us believe that such a theory with the desired form of $Z$ as in \eqref{formpart} would exist and if it does, how different would that be from pure gravity? The answer comes from the analogous computation done in case of AdS background \cite{Gaberdiel:2010ar,Gaberdiel:2010pz}. This is in the sense that if one minimally couples higher spin fields with pure gravity, the linearized higher spin action for a given spin would result into a 1-loop partition function structurally similar to that of the spin-2 case. Finally one would take product of partition functions of all spin interactions upto highest spin in the truncated tower of spins. This qualitative argument therefore leads us to speculate that we need at least to include spins=3 and 4 in order to get a convergent answer in addition to spin 2. That would correspond to $l=3$ in light of \eqref{formpart}.

An ideal case would be to perform the analysis for arbitrary higher spin and then see the critical spin for which divergence is removed. However that would be accompanied with calculational complexity, which is not relevant to the present goal. We will however perform in its full detail the case of spin 3 in the following section. This computation will indicate the forms of all higher spins.
\section{Enter the higher spins}
We now consider a finite tower of higher spins $s=3,4, \cdots N$ present in the dynamics, all minimally coupled to background spin-2 fields, linearized and massless. We closely follow the analysis of \cite{Gaberdiel:2010ar}. The case of spin 3 is most tractable as far as computational complication is concerned. We would furnish the one-loop partition function for spin-3 interctions coupled to gravity in what follows. The case of arbitrary spins will then be presented omitting the detailed steps.

The background field $g_{\mu \nu}$ will be considered as describing Euclidean de Sitter in the static patch. The Riemann curvature tensor is defined via the metric compatible connection $\nabla$ and a probe vector field $A^{\mu}$:
$$ [\nabla_{\mu}, \nabla_{\nu}] A^{\rho} = R^{ \rho}{}_{\sigma \mu \nu} A^{\sigma}.$$
Since this has constant Ricci scalar and the Weyl tensor vanishes in 3 dimension,
 $$R^{ \rho}{}_{\sigma \mu \nu} = (\delta^{\rho}{}_{\mu}\, g_{\sigma \nu} - \delta^{\rho}{}_{\nu}\, g_{\sigma \mu}),~~~R_{\mu \nu} = 2 g_{\mu \nu}$$
\subsection{The spin-3 determinant}
A massless spin-3 field $\phi_{\mu_1 \mu_2 \mu_3}$ (completely symmetrized) that can be minimally coupled to pure gravity in 3 space-time dimensions is governed by the following linearized Fronsdal \cite{Fronsdal:1978rb} action in addition to \eqref{sp2} 
\bea \label{sp3}
S_{\mathrm{spin-}3}[\phi] = \int \sqrt{g} \,\phi^{\mu_1 \mu_2 \mu_3} \left( \hat{\mathcal{F}}_{\mu_1 \mu_2 \mu_3} - \frac{1}{2} g_{( \mu_1 \mu_2} \hat{\mathcal{F}} _{ \mu _3) \nu}{}{}^{\nu}\right) d^{3} x \eea
where, 
$$\hat{\mathcal{F}}_{\mu_1 \mu_2 \mu_3} = \Delta \phi_{\mu_1 \mu_2 \mu_3} - \nabla _{(\mu_1} \nabla ^{ \lambda} \phi_{ \mu_2 \mu_3) \lambda}+\frac{1}{2} \nabla_{(\mu_1} \nabla_{\mu_2} \phi _{\mu_3 )\lambda}{}{}^\lambda + 2 g_{(\mu_1 \mu_2} \phi_{\mu_3) \lambda}{}{}^\lambda .$$
This linearized version of the theory suffices our purpose of computing functional integral upto 1-loop level. Here $\Delta$ is the Laplacian associated with the connection $ \nabla $ compatible with the background $g$. As per convention for the symmetrizer used above, we do not use any combinatorial factor. But for terms like $\nabla_{(\mu_1} \phi _{\mu_2 \mu_3  \mu_4)}$ containing already symmetric tensor $ \phi$, we do not consider any more terms coming from permutations within $ \mu_2, \mu_3, \mu_4$.

While going to construct the 1-loop determinant for the above theory, we first note that there is a gauge invariance which is easy to see at the linearized level:
$$ \delta \phi_{\mu_1 \mu_2 \mu_3} = \nabla_{(\mu_1} \xi_{\mu_2 \mu_3)} ,$$
the gauge parameter $\xi$ is a symmetric traceless rank-2 tensor. With this observation $\phi$ can be decomposed as:
\bea \label{split}
\phi_{\mu_1 \mu_2 \mu_3} = \phi^{\mathrm{TT}}_{\mu_1 \mu_2 \mu_3} + g_{(\mu_1 \mu_2} \psi_{\mu_3)} + \nabla_{(\mu_1} \xi_{\mu_2 \mu_3)}
\eea
where the first term is the transverse (read divergence-less), traceless part. The vector or spin-1 part $\psi_{\mu}$ encodes the trace information and $\xi$ is the gauge component. One can further decompose the vector part into transverse and gradient components:
\bea \label{gaugesplit} \psi _{\mu} = \psi^{\mathrm{T}}_{\mu} + \nabla _{\mu} \chi \eea
The idea for computing the 1-loop determinant essentially follows \cite{Gaberdiel:2010xv}. We do not fix a specific gauge here; but instead extract the gauge part explicitly and use the gauge invariance of the action to get rid of the gauge redundancy. For the case of de Sitter quantum gravity this method of deriving the 1-loop determinant was employed first in \cite{Vassilevich:1992rk, Vassilevich:1995wh}. Similar gauge-independent quantization is well-understood in flat space quantum field theory, see for example \cite{Bhattacharjee:2013kja},\cite{Bhattacharjee:2010ie}, \cite{Chernodub:2008rz} and references therein for recent developments. Using the decompositions \eqref{split} and \eqref{gaugesplit} and appealing to the gauge invariance (read independence of $ \xi$) of \eqref{sp3} we have:
\bea
\label{splits3}
S_{\mathrm{spin-}3}[\phi] = \int \sqrt{g} \, [\phi^{\mathrm{TT}}_{\mu_1 \mu_2 \mu_3} (- \Delta) \phi^{\mathrm{TT}\, \mu_1 \mu_2 \mu_3} + 36 \psi ^{\mathrm{T}\, \mu } (- \Delta - 7) \psi^{\mathrm{T}}_{\mu} + \frac{81}{2} \chi (- \Delta-8)(-\Delta) \chi]\eea
Note that the decomposition \eqref{split} and \eqref{gaugesplit} pays off by completely decoupling the $ \phi^{ \mathrm{TT}}, \psi^{\mathrm{T}} \mbox{ and } \chi$ modes. This leads to readily reading off the 1-loop functional determinant from the linearized action \eqref{splits3}:
\bea \label{partnaive}
Z^{[\mathrm{1-loop}]}_{(3)} = \dfrac{1}{\mathrm{Vol[gauge\,group]}}\int\, [D \phi] \, e^{ - S_{\mathrm{spin-}3}[\phi] }
\eea

However the decomposition of the variables $ \phi  \rightarrow (\phi^{\mathrm{TT}}, \psi^{\mathrm{T}}, \chi, \xi )$ would require a careful calculation of the Jacobian involved in the transformation of the functional measure. It is now instructive to decompose the gauge mode $\xi_{\mu \nu}$ in its different irreducible helicity components as well:
\bea \label{splitxi}
\xi_{\mu \nu } = \xi^{\mathrm{TT}}_{\mu \nu } +  \nabla_{(\mu} \sigma _{\nu)} - \frac{2}{3}g_{\mu \nu} \nabla^{\rho} \sigma_{\rho}
\eea
The first term is the usual transverse-traceless part. The last one term in the longitudinal part is kept in view of overall tracelessness of $ \xi$. But the vector gauge mode $ \sigma$ can further be decomposed into a transverse vector ($ \sigma^{\mathrm{T}}_{\mu}$) and a longitudinal scalar ($ \omega$), such that:
\bea \label{xidecomp}
\xi_{\mu \nu } = \xi^{\mathrm{TT}}_{\mu \nu } +  \nabla_{(\mu} \sigma^{\mathrm{T}} _{\nu)} + 2 (\nabla_{\mu} \nabla_{\nu} - \frac{1}{3} g_{\mu \nu}\nabla^2)\, \omega
\eea

One should actually now concentrate on the Jacobian $J$ for the full decomposition (transformation) $ \phi  \rightarrow (\phi^{\mathrm{TT}}, \psi^{\mathrm{T}}, \chi, \xi^{\mathrm{TT}}, \sigma^{\mathrm{T}}, \omega )$ where 
\bea \label{measuresplit}
[D \phi] \mapsto [D \phi^{\mathrm{TT}}] [D \psi ^{\mathrm{T}}] [D \chi] [D \xi^{\mathrm{TT}}] [D \sigma ^{\mathrm{T}}] [D \omega] = [D \phi] J^{-1}.\eea
In order to do so we choose the following convention for functional measures:
\bea
\int \, [DA] \, \exp(- \int d^3 x \sqrt{g}\, A^{\ast}\, A_{\ast}) =1 .
\eea
Here for any field $A$, $\ast$ denotes arbitrary index structure making the integrand invariant under all possible local and global symmetries. This would determine the Jacobian $J$ through:
\bea \label{defJ}
J \,\int  [D \phi^{\mathrm{TT}}] [D \psi ^{\mathrm{T}}] [D \chi] [D \xi^{\mathrm{TT}}] [D \sigma ^{\mathrm{T}}] [D \omega] \, \exp(- \int d^3 x \sqrt{g}\, \phi^{\mu_1 \mu_2\mu_3}\, \phi_{\mu_1 \mu_2\mu_3} ) =1
\eea
We note here that the change of variables, which can be viewed as a coordinate transformations in field space is linear. That is what guarantees that the Jacobian is independent of the fields and was be pulled out of functional integral. The expression of $J$ is carried out in detail in the Appendix and the answer is \eqref{jacob}. Thanks to the split \eqref{measuresplit} in measure, the non-gauge (physical) modes in \eqref{partnaive} can be easily integrated out leaving the decoupled gauge part and the Jacobian:
\bea \label{part}
Z^{[\mathrm{1-loop}]}_{(3)} &=& \dfrac{J}{\mathrm{Vol[gauge\,group]}} \int [D \phi^{\mathrm{TT}}] [D \psi ^{\mathrm{T}}] [D \chi][D \xi^{\mathrm{TT}}] [D \sigma ^{\mathrm{T}}] [D \omega]  \non \\
        && \exp \left(-\langle \phi^{\mathrm{TT}}, (- \Delta) \phi^{\mathrm{TT}}\rangle  -36 \langle\psi ^{\mathrm{T}},(-\Delta -7)\psi ^{\mathrm{T}}\rangle -\frac{81}{2}\langle \chi,  (- \Delta-8)(-\Delta)\chi \rangle\right) \non \\
        &=& [\mathrm{det}(-\Delta)^{\mathrm{TT}}_{(3)}\, \mathrm{det}(-\Delta- 7)^{\mathrm{T}}_{(1)}\mathrm{det}\{(-\Delta)_{(0)}(-\Delta- 8)_{(0)}\}]^{-1/2} \non \\ & \times &  \dfrac{J \displaystyle \int [D \xi^{\mathrm{TT}}] [D \sigma ^{\mathrm{T}}] [D \omega]}{\mathrm{Vol[gauge\,group]}}.
\eea
As usual, here $ \mathrm{det}(-\Delta)_{(3)}^{\mathrm{TT}} $ stands for the determinant of the Laplacian operator corresponding to transverse, traceless spin-3 modes only. Similar notation applies for the other determinants as well. Now, the formal definition of the gauge group volume is:
$$\mathrm{Vol[gauge\,group]} = \int [D \xi].$$
We therefore need to evaluate another Jacobian 
\bea \label{jac2}\tilde{J}=\dfrac{[D\xi]}{[D \xi^{\mathrm{TT}}] [D \sigma ^{\mathrm{T}}][D \omega]} \eea
in order to find the ratio in the last term of \eqref{part}. This is done explicitly in the Appendix and the end result is \eqref{jac2eval}. With this definition and the ratio of $J$ and $ \tilde{J}$ from \eqref{ratio} we arrive at a much simplified formula, involving only spin-3 and spin-2 determinants:
\bea \label{spin3det}
Z^{[\mathrm{1-loop}]}_{(3)} &=& [\mathrm{det}(-\Delta)^{\mathrm{TT}}_{(3)}\, \mathrm{det}(-\Delta- 7)^{\mathrm{T}}_{(1)}\mathrm{det}\{(-\Delta)_{(0)}(-\Delta- 8)_{(0)}\}]^{-1/2} \dfrac{J}{\tilde{J}} \non \\
&=& \left[ \dfrac{\mathrm{det}(-\Delta- 6)^{\mathrm{TT}}_{(2)}}{\mathrm{det}(-\Delta)^{\mathrm{TT}}_{(3)}}\right]^{1/2}
\eea
\subsection{$Z^{[\mathrm{1-loop}]}_{(3)}$ on $L(p,q)$ from heat Kernel}
Now that we have partition function in a more tractable form \eqref{spin3det}, the following steps will be evaluation of the determinants. As a first step toward this, let us remind ourselves that we need to calculate spectra of operators, which are Laplacian, augmented by some constant at most,:
\begin{itemize}
\item on static patch of thermal de Sitter, ie $S^3$ endowed with smooth metric and its smooth quotients $L(p,q)$ defined in the Introduction,
\item acting on the space of `square integrable' tangent space tensors (with particular property of being symmetric, traceless and transverse).
\end{itemize}
A great body of authoritative work on harmonic analysis on Lens spaces, accessible from physicists' perspective has been accomplished by JS Dowker. See for example \cite{Dowker:2004nh}. Another extremely useful resource for these results is \cite{David:2009xg}, where a group theoretic route for the harmonic analysis on $S^3$ and its quotients was taken. They used the fact that $S^3$ can be viewed as the homogeneous space $G/H$ with $G = SU(2 ) \times SU(2)$ and $H =SU(2)$ with $H$ acting `diagonally' on $G$. The spin-s tensors arranged in eigenbasis of the spin-s Laplacian are then sections of the associated vector bundle $\pi: E_s \rightarrow G/H$ of the principal bundle $ \pi^{\prime} : G \rightarrow G/H$. We will be using their results in our analysis.

\vspace{0.4cm}
\textbf{\underline{Scheme of the computation}}

Since we are talking about elliptic operators on compact-closed manifolds, the spectra are obviously discrete. If the $n^{\mathrm{th}}$ eigenvalue of the operator $O$ is $\lambda_n$ (say, positive definite) 
with degeneracy $d_n$, then the determinant we wish to evaluate is formally:
\bea \label{Odet}
 \mathrm{det}(-O)=\prod_n (-\lambda_n)^{d_n}.
\eea
A convenient way to encode the information of the spectrum is through the the non-local object, the heat-Kernel:
\bea \label{hk}
K(x,y;t) = \sum_n \varphi_n(x) \varphi^{\ast}_n(y) e^{- \lambda_n t}.
\eea
Where $\varphi_n(x)$ are the normalized eigenfunctions of the operator $O$. By taking logarithm of \eqref{Odet}, it is now fairly straightforward to see that
\bea \label{logOdet}
\ln \, \mathrm{det}(-O) = \mathrm{tr} \ln (-O) &=&  -\int_{0}^{\infty} \frac{dt}{t} \int \sqrt{g}\, d^3 x \,K(x,x;t) \nonumber\\
&=:& -\int_{0}^{\infty} \frac{dt}{t} K(t)
\eea
where we have used the completeness of the eigenfunctions $ \varphi _n (x)$ to manipulate the coincident heat Kernel $ K(x,x;t)$. Another small observation that would be relevant for our case is that for a c-number $a$, the transformation $ O \rightarrow O + a$ does not alter the spectrum except giving the eigenvalues a constant shift by $a$. However the determinant \eqref{logOdet} changes according to:
\bea \label{logOdeta}
 \ln \, \mathrm{det}(-O-a) = -\int_{0}^{\infty} \frac{dt}{t} K(t) \, e^{ a t}
\eea
Depending upon the nature of the spectrum this above integral may diverge and therefore would need a regularization. For example the following regularization, as we will see, is apt for the problem at hand:
\bea
\mbox{define }~~  \zeta(z;a) &=& \frac{1}{\Gamma(z)} \int_{0}^{\infty} \frac{dt}{t^{1-z}} \int \sqrt{g}\, d^3 x \,K(t) e^{ a t} \non \\
&=& \sum_n \frac{d_n}{(\lambda_n +a)^z}, \non  \\
\mbox{such that }~~ - \lim_{z \rightarrow 0} \frac{d}{dz} \zeta (z;a) &=& \ln \, \mathrm{det}(-O-a)
\eea
would give us the regulated answer for the determinant, given the analytic property of these zeta functions on the complex plain.

\vspace{0.4cm}
\textbf{\underline{Details of the computation}}

To appreciate the feasibility of the computation, let us first see the explicit spectrum for the simplest case of $S^3$, as presented in \cite{David:2009xg}. For the Laplacian, $(\Delta)^{\mathrm{TT}}_{(s)}$ that act on the $L^2$ space of spin-$s$ symmetric transverse-traceless tensors, the eigenvalues and degeneracies of states are given by:
\bea \label{spectrum}
\lambda_n &=& (s+n)(s+n+2)- s ,~~~~ n \geq 0 \non \\
d_n &=& (2 - \delta_{s,0} ) (n + 1) (n + 2s + 1).
\eea
One observes the absence of any zero mode in the spectrum of the operators relevant to our case \eqref{spin3det}, ie for spins $s=1,3$. The integral of coincident heat Kernel \eqref{logOdet} for $(- \Delta)^{\mathrm{TT}}_{(s)}$ on $S^3$ background can be conveniently expressed as
\bea
K^{(s)}(t) = (2- \delta_{s,0}) \sum _{n=s+1}^{ \infty} (n^{2} - s^2) e^{( -n^2 +s+1)t}
\eea
\vspace{0.5cm}

The same scheme follows for these operators on general Lens spaces. The difference occurs in the spectrum, which becomes more involved to analyse. Thankfully that was also calculated in \cite{David:2009xg}. As we know Lens spaces are obtained from $S^3 \sim SU(2)$ by quotienting by a discrete group. Keeping this in mind, heat kernel for spin-s Laplacians were derived by the method of images starting from the $S^3$ result to find:
\bea \label{hklens}
K^{(s)}(t) &=&\frac{1}{p}\sum_{n=0}^{\infty} \sum_{m\in \mathbb{Z}_{p}}\left[\chi_{\frac{n}{2}}\left( m \tau\right)\chi_{\frac{n}{2}+s}\left( m \bar{\tau}\right) + \chi_{\frac{n}{2}+s}\left( m \tau\right)\chi_{\frac{n}{2}}\left( m \bar{\tau}\right)\right] e^{ -\lambda_n t}\non\\
\mbox{with }~\chi_l(\theta) &=& \frac{\sin((2l+1)\theta/2)}{\sin(\theta/2)} ~~~\mbox{being\, SU(2)\, characters.}
\eea
$\lambda_n$ are the same as in the case of $S^3$, ie in \eqref{spectrum}. Here the parameters $\tau, \bar{\tau}$ are defined
\footnote{These were introduced in a form that suggests a resemblance with the boundary modular parameter of `upper-half' Hyperbolic space $H_{3}^{+}$. Actually $H_{3}^{+}$ is the Euclidean version of AdS${}_3$ (or thermal AdS), whose boundary is a 2-torus described by the modular parameter $\tau$. In this sense, the real $\tau$ defined above can be thought of as the analytically continued version of the Eucledian AdS${}_3$ modular parameter. This was given a firm ground in \cite{David:2009xg} by establishing that the corresponding heat Kernels do match by this analytic continuation. However an essential distinction remains in the fact that $\tau$ can only take distinct rational (upto a $\pi$ factor) values for Lens pace family, each fixed $\tau$ designating a particular Lens space. On the other hand the modular parameter takes value in a connected sub-set of complex upper half plane. Physically this is equivalent to the observation that only discrete values of temperature and angular potential describe static patches of de Sitter while a continuous set of values can be ascribed to the corresponding quantities in thermal AdS${}_3$.} as:
$$ \tau = \frac{2 \pi (q-1) }{p},\, \bar{\tau} = \frac{2 \pi (q+1) }{p}.$$ From \eqref{hklens} it is clear that all the quotients $S^3/ \mathbb{Z}_p$ share the same eigenvalues $\lambda_n$ of $ \Delta_{(s)}$ for a given spin as those of $S^3$. It is the distribution or the degeneracy of eigenvalues, that is different for each quotient.

With all necessary ingredients at hand let us write the spin-3 1-loop partition function (rather its logarithm):
\bea \label{logZ3}
\ln Z^{[\mathrm{1-loop}]}_{(3)} &=& \frac{1}{2} \left[ \ln \left( - \Delta^{\mathrm{T}}_{(1)} -6 \right) - \ln \left( - \Delta^{\mathrm{T}}_{(3)} \right) \right] \non \\
&=& \frac{1}{2} \frac{d}{dz} \left[ \frac{1}{ \Gamma(z)} \int _{0}^{ \infty} t^{z-1} \left( K^{(3)}(t) - K^{(1)}(t) e^{6t}\right)\right]_{z=0}
\eea 
We now use \eqref{hklens}, perform the $t$ integral and some manipulations involving the summand (shifting the $n$ sum) to get the last line to be equal to:
\bea \label{exprn}
\frac{1}{p}  \sum_{n=4}^{\infty} \sum_{m \in \mathbb{Z}_p}&& \Bigg[-\left( \cos\left(\frac{6 m \pi q} {p}\right) \cos\left(\frac{2 m n \pi }{p}\right)  - \cos\left(\frac{6 m \pi }{p}\right) 
\cos\left(\frac{2 m n \pi q}{p}\right)\right)\, \ln(n^2 -4)  \non \\ &+& 
\left( \cos\left(\frac{4 m \pi q}{p}\right) \cos\left(\frac{2 m n \pi }{p}\right) - 
\cos\left(\frac{4 m \pi }{p}\right) \cos\left(\frac{2 m n \pi q}{p}\right) \right) \, \ln 
(n^2 -9)\Bigg] \non \\
& \times &
\frac{1}{\left( \cos\left(\frac{2m \pi }{p}\right)-\cos\left(\frac{2m \pi q }{p}\right)\right)}
\eea
The above sum of course is divergent due to the terms $\sim \ln (n^2)$ and one needs to be regulate it through zeta function regularization. 
Let us replace the above sum by a suitable zeta function. 
We first introduce a couple of `oscillatory zeta' functions:
\bea \label{osczet}
\tilde{\zeta}_1 (\theta;z) = \sum_{n=4}^{\infty}\cos \l n \theta \r \left[ \l n-2\r ^{-z} +  \l n+2\r ^{-z}\right] \non \\
\tilde{\zeta}_2 (\theta;z) = \sum_{n=4}^{\infty}\cos \l n \theta \r \left[ \l n-3\r ^{-z} +  \l n+3\r ^{-z}\right] \non.
\eea
The idea is to finally replace the log of the 1-loop partition function \eqref{exprn} in terms of these zeta functions as:
\bea \label{exprn2}
\frac{1}{p} \sum_{m \in \mathbb{Z}_p} \frac{1}{\cos\left(m\theta_2\right)-\cos\left(m\theta_1\right)}&&\frac{d}{dz} \Big[ \cos(3 m\theta_1) \zeta_1 (m\theta_2;z)- \cos(3m \theta_2) \zeta_1 (m\theta_1;z) \non \\
&&-\cos(2 m\theta_1)\zeta_2 (m\theta_2;z)+\cos(2 m\theta_2)\zeta_2 (m\theta_1;z)
\Big]_{z=0} 
\eea
where $\theta_1 = \frac{\tau + \bar{\tau}}{2} =\frac{2 \pi q}{p} , ~~\theta_2 =  \frac{-\tau + \bar{\tau}}{2} =\frac{2 \pi }{p}$. Next one defines another `oscillatory zeta':
$$ \tilde{\zeta}_0(\theta;z) = \sum_{n=1}^{\infty} \frac{\cos \l n \theta \r }{n^z}$$ which is related to the previously defined 2 by simple trigonometric identities:
\bea \label{morezeta}
\tilde{\zeta}_1 (\theta; z) &=& 2 \cos\l 2 \theta\r \tilde{\zeta}_0 (\theta; z) -\sum_{n=2}^{5} \frac{\cos\l(n-2) \theta\r}{n^z} + \left( \mbox{ terms independent of }z\right) \non \\
\tilde{\zeta}_2 (\theta; z) &=& 2 \cos\l 3 \theta\r \tilde{\zeta}_0 (\theta; z) -\sum_{n=2}^{6} \frac{\cos\l(n-3) \theta\r}{n^z} +\left( \mbox{ terms independent of }z\right) 
\eea
If we plug \eqref{morezeta} into \eqref{exprn2}, a long yet straightforward set of manipulations involving trigonometric identities leads one to:
\bea \label{semifexprn}
&& \ln Z^{[\mathrm{1-loop}]}_{(3)} \non\\
&=& -\frac{2}{p} \frac{d}{dz}\sum_{m \in \mathbb{Z}_p}  \l 1+ \cos(m \tau)+ \cos(m \bar{\tau})+ \cos(2 m \tau)+ \cos(2 m \bar{\tau})\r  \l \tilde{\zeta}_0(m \theta_1 ;z) + \tilde{\zeta}_0(m \theta_2 ;z)\r \Bigg|_{z=0}  \non\\
&+& \frac{d}{dz} \Bigg[ 6^{-z} + 2^{1-z} \delta _{p,2} + 3^{-z} \delta_{q,2} + (6^{-z}+2^{-z})\delta_{q,1} +(6^{-z}+4^{-z})(\delta_{q,1}+ \delta_{2q,p+2}) \\
 &+& (6^{-z}+2^{-z})\delta_{q,p-1}+ (6^{-z}+4^{-z}) (\delta_{q,p-1}+ \delta_{2q,p-2})+3^{-z}(\delta_{q,p-2}+\delta_{2q,p-1}+\delta_{2q,p+1})\Big]_{z=0} \non
\eea
Although the above formula does not reveal much of physics at first sight, we can gather some quick insights from it. If we concentrate on the first part of the expression above, ie the $m$ summand, the terms $\cos(2 m \tau), \cos(2 m \bar{\tau})$ are the first genuine contributions from the spin-3 dynamics. The three terms preceding it are same as those for pure spin-2 fields at 1-loop of perturbation theory. As we will see later that this is a generic feature of any higher spin tower, ie any spin ($>$2) will have all of the lower spins in the tower contributing in the quantum theory. Another point to note is that the Kronecker delta terms appearing in the later part of the formula \eqref{semifexprn} do not contribute significantly to the partition function except for providing overall numerical constant, independent of at some special values of $p,q$. 

Further simplification of the first part (ie, the $\mathbb{Z}_p$ sum) of the above formula requires few more steps, which we explain elaborately in Appendix \ref{app2} picking up a sample term \\
$-\frac{2}{p} \frac{d}{dz}\displaystyle{\sum_{m \in \mathbb{Z}_p}} \cos( m \tau) \tilde{\zeta}_{0}(m \theta_1 ; z)$. It is to be noted that for convenience, we would be dealing with the special cases for the pair $(q,p)$:
\bea \label{special}q = (\pm 1, \pm 2)\mbox{ mod }p ~~\mbox{ and }~~ 2q= (\pm 1, \pm 2) \mbox{ mod } p
\eea separately. Except for these special cases , the final form of the 1-loop partition function for spin-3 is:
\bea \label{det2}
 Z^{[\mathrm{1-loop}]}_{(3)} = \frac{2^9 \pi ^2}{3 p^2} \, \prod_{r=1}^{2}\prod_{\pm} \sin \left( r \pi  \frac{q \pm 1}{p} \right)\sin \left( r \pi  \frac{q^{ \ast } \pm 1}{p} \right),
\eea
where $q^{\ast}$ is a positive integer modulo $p$, such that $q q^{\ast} = 1~\mathrm{mod}~ p$.

\vspace{0.4cm}

\textbf{\underline{The special cases}}

The above analysis does not hold for the special cases \eqref{special} as mentioned above. For example, for the case $q= \pm 1 \mbox{ mod } \, p$, (except for $p=2, q=1$) the similar analysis as outlined before, yields:
\bea \label{det_sp1}
 Z^{[\mathrm{1-loop}]}_{(3)} \sim \frac{1}{p^8} \sin ^2 \left(   \frac{2 \pi}{p} \right)\sin ^2 \left(   \frac{4 \pi}{p} \right) \non
\eea
up to numerical factors. The other cases ie, $ q = \pm 2 \mbox{ mod }\, p$ and $2q = (\pm 1, \pm 2) \mbox{ mod }\,p$ has the same feature of $p^{-8} \times \mbox{ trigonometric function }(p)$. This essentially allows us to not care about these special cases, as long as we are concerned about the convergence of the partition functions over all Lens spaces.


\subsection{Zero modes}
Up to this point we have not talked about possible zero modes that could inflict our analysis of the determinant calculation. Here we wish to shed light on this aspect. There are 2 operators in \eqref{spin3det} of concern, whose zero modes can result into meaningless answers. By inspection we notice that rank-2 symmetric, traceless, transverse tensors $\xi_{\mu \nu}$ satisfying
\bea \label{Killing} \nabla_{(\mu} \xi_{\nu \rho)} = 0 \eea
ie, the rank-2 Killing tensors indeed form the kernel of the `ghost operator' $(\Delta + 6)^{\mathrm{TT}}_{(2)}$. This can be easily verified by taking divergence of the left hand side of \eqref{Killing} and using properties of the constant positive curvature space with Riemann scalar $R=6$.

According to \cite{Thompson:1986jmp}, the Killing tensors of constant curvature Riemann manifold can be constructed as symmetrized tensor products of the Killing vector fields of the manifold. Rank-2 Killing tensors in 3 dimensional constant curvature manifold form a vector space of dimension 20 \cite{Thompson:1986jmp}. The traceless condition reduces this number to 10 \cite{Giombi:2013yva}. However a general Lens space $L(p,q)$ is not a maximally symmetric space and only admits two globally defined Killing vector fields. They Lie commute, generating the group $U(1) \times U(1)$. We will see in a trivial analysis that the traceless, rank-2 symmetric tensor space constructed out if it, is also of dimension $n=2$.

Let $\xi^{A}_{\mu}$ with $A=1,2$ be a convenient basis on the 2-dimensional real vector space $V_{(1)} =  \mathfrak{u}(1) \oplus \mathfrak{u}(1)$ of the Killing vectors. Using the above mentioned result from \cite{Thompson:1986jmp}, we see a general rank-2 Killing tensor must be of the form:
\bea \label{KT}
\xi_{ \mu \nu } = C_{AB} \xi^{A}_{\mu} \xi^{B}_{\nu}
\eea
where $C _{AB}$ is a 2-dimensional real symmetric constant matrix. It has 3 independent components. Demanding $\xi_{\mu \nu}$ to be traceless, imposes the constraint $C_{AB} \xi^{A}_{\mu} \xi^{B \mu} =0$. This reduces \footnote{Another way of seeing this is as follows. The vector space $V_{(1)}$ furnish a 2-dimensional representation of the group $U(1)\times U(1)$. We impose a symmetric invariant bilinear form $\delta^{AB}$ on $V_{(1)}$. Note such a structure does not naturally come from the metric defined by the Lie algebra of the Killing vector fields, which is trivially zero because of the algebra is Abelian. As elements of $V_{(1)}$, the $\xi^A$ form a complete set, such that:
\bea \label{assumbil}
g^{\mu \nu} \xi^{A}_{\mu} \xi ^{B}_{\nu} = \frac{1}{2} \delta^{AB} \delta_{CD} \xi^{C}_{\mu} \xi ^{D \, \mu}.\non 
\eea
This implies that:
\bea
g^{\mu \nu} \xi_{\mu \nu} = \mathrm{Tr(C)} \delta_{AB} \xi_{\mu}^A \xi_{nu}^B
\eea
$\xi_{\mu \nu}$ as in \eqref{KT} can be traceless if and only if trace of $C_{AB}$ with respect to $\delta^{AB}$ vanishes. This reduces the number of real independent components of $C$ to 2.} the number of independent components of $C$ to 2. Hence the space of symmetric, traceless, rank-2 Killing tensors  $V_{(2)} \subset V_{(1)} \times V_{(1)}$ is 2-dimensional
The presence of these zero modes indicate that the sectioning of the space of spin-3 fields into gauge modes and gauge invariant ones was not completely specified by \eqref{split}. There were gauge redundancies which were over-looked by this process generating trivial gauge transformations. These are the Killing tensors discussed right now. From this point of view we must correctly calculate the volume of the gauge group appearing in \eqref{partnaive} by changing the measure. That analysis is extremely well explained for $S^3$ in \cite{Giombi:2013yva}. 
This analysis  and that of \cite{Castro:2011xb} suggests that we include a factor 
\bea \label{zm}
D_{\mathrm{zm}} = k^{-1} \left(\mathrm{Vol} (L(p,q))\right)^{-n/2} = \dfrac{p}{2 \pi ^2 k}, \eea which corrects the problem of measure on space of gauge modes, in \eqref{det2}. One may note that this factor contains an inverse power of $k$. Therefore $D_{\mathrm{zm}}$ multiplied to the 1-loop determinant result will give rise to a term $\exp(-\log(k))$. Now the general structure of perturbative partition function should be of the form:
$$Z \sim \exp \left(-k S_0 + S_1 + k^{-1}S_2 + k^{-2} S_3+ \cdots \right)$$
where $S_n$ is the $n$th loop $k$ independent effective action. This indicates that the zero mode contribution is of non-perturbative nature and corresponds to an instanton. This was also a feature of the spin-2 case \eqref{ls1}.

\subsection{Still diverging as expected}
The total partition function up to 1-loop and with spin-3 dynamics, for a particular Lens space should be:
$$ D_{\mathrm{zm}} Z^{[\mathrm{tree}]}_{(2)} \, Z^{[\mathrm{1-loop}]}_{(2)}\, Z^{[\mathrm{tree}]}_{(3)}\, Z^{[\mathrm{1-loop}]}_{(3)}$$
But the spin-3 dynamics is considered fluctuations over trivial vacuum. This gives essentially no contribution in terms of $Z^{[\mathrm{tree}]}_{(3)}$. We therefore use \eqref{ls1}, \eqref{det2} and \eqref{zm} to get 1-loop partition function of spin-3 gravity for a given pair $p,q$:
\bea \label{sp3_pq}
Z(p,q) = \frac{C_1 e^{2 \pi k /p}}{k^2 \, p^2} \prod_{\pm} \sin ^2 \left(\pi  \frac{q \pm 1}{p} \right)\sin ^2 \left(  \pi  \frac{q^{ \ast } \pm 1}{p} \right)\, \sin  \left(2\pi  \frac{q \pm 1}{p} \right)\sin  \left(2  \pi  \frac{q^{ \ast } \pm 1}{p} \right)
\eea
Here we have included the irrelevant numerical factors in $C_1$. Now a comparison with the suggestive form \eqref{formpart} reveals that $l=2$ for \eqref{sp3_pq}, which clearly fails to make the total partition function, when summed over all $p,q$, divergent.
\section{Spin$\ge 4$ and Convergence}
\subsection{Convergence for spin-4}
The case of spin-4 and higher can be seen by an almost similar analysis. The basis dynamical variables are now rank-4 symmetric tensors with gauge transformation:
$$ \delta \phi_{\mu_1 \dots \mu_4} = \nabla_{(\mu_1} \xi_{\mu_2 \mu_3 \mu_4)}$$
which keeps the linearized Fronsdal action invariant. The gauge parameters $\xi$ here are traceless symmetric rank-3 tensors. The linearized action and the same tricks of splitting $\phi$ into physical and gauge modes, as done before gives one the 1-loop partition function for the spin-4 field:
\bea \label{spin4det}
Z^{[\mathrm{1-loop}]}_{(4)}= \left[ \dfrac{\mathrm{det}(-\Delta - 12)^{\mathrm{TT}}_{(3)}}{\mathrm{det}(-\Delta - 4)^{\mathrm{TT}}_{(4)}}\right]^{1/2}.
\eea
Again a very long evaluation of it, which exactly analogous to the earlier computation gives
\bea \label{det4}
Z^{[\mathrm{1-loop}]}_{(4)}= \frac{C_2}{p^2} \, \prod_{r=1}^{3}\prod_{\pm} \sin \left( r \pi  \frac{q \pm 1}{p} \right)\sin \left( r \pi  \frac{q^{ \ast } \pm 1}{p} \right)
\eea
except for the special values $m\,q = (\pm 1, \pm2, \pm 3) \, \mathrm{mod}\,p$ with $m=1,2,3$. For these values $Z^{[\mathrm{1-loop}]}_{(4)} \sim \frac{1}{p^{12}} \times \mbox{ Trigonometric functions}(p)$ and therefore are harmless in our convergence analysis. 

As in the case of spin-3 we will have to consider zero modes here too. They form the Kernel of the operator in the numerator :$ (-\Delta - 12)^{\mathrm{TT}}_{(3)} $, which are traceless rank-3 Killing tensors. A generic form of such a tensor should be of the form
$$ \xi_{\mu \nu \rho} = C_{ABC} \xi ^A_{\mu} \xi ^B_{\nu} \xi ^C_{\rho}$$ 
with $C_{ABC}$ is symmetric and $\xi^A$ form the 2-dimensional basis for Killing vectors. Arguments as presented for the spin-3 case now enforces that $C$ should be ripped of all its traces. This remarkably makes independent components of $C_{ABC}$ to be 2 again. Hence traceless rank-3 Killing tensors form a 2 dimensional vector space here too. This means that the zero mode correction will be exactly same as \eqref{zm}.

Collecting all the factors (tree level part for spin-4 is also taken to be trivial), we find upto spin-4, the partition function on $L(p,q)$
\bea \label{spin-4}
Z(p,q) &=& \frac{C_3 \, e^{2 \pi k/p}}{(k\,p)^3} \prod_{\pm}   \sin \left(  \pi  \frac{q \pm 1}{p} \right)\sin \left(  \pi  \frac{q^{ \ast } \pm 1}{p} \right)  \l \prod_{r=1,2} \sin \left( r \pi  \frac{q \pm 1}{p} \right)\sin \left( r \pi  \frac{q^{ \ast } \pm 1}{p} \right) \r \non \\
&& \times \l \prod_{t=1}^3 \sin \left(t  \pi  \frac{q \pm 1}{p} \right)\sin \left( t \pi  \frac{q^{ \ast } \pm 1}{p} \right) \r
\eea
Let us examine how it is beneficial in convergence of a sum like \eqref{def}. Note that the special cases of $m\, q = (\pm 1,2,\pm 3 ) \, \mbox{ mod }p $  with $m=1,2,3 $ are of slightly different form than \eqref{sp3_pq}, as their $p$ suppression is of power greater than 2, as already mentioned above. However we will make a harmless assumption that the form \eqref{sp3_pq} holds for all $q, p$. If we can show that the sum of \eqref{sp3_pq} over all Lens spaces is meaningful, it surely would be so for the one with careful analysis of the special values taken care of. 

At this juncture we will explicitly show that the divergence from $ \zeta(1)$ that appeared from $p^{-l}$ suppression of the partition function for $l<3$ can be removed with spin-4 dynamics. Although this can be seen from the qualitatively based on the argument following \eqref{formpart}. We still need to show that there are no more divergence wreaking validity of \eqref{spin-4}. For this part, we present here a simple comparison test. 

Summing over all Lens spaces would give us \footnote{There is a slight caveat is there in the sum. Note that the two Lens spaces $L(p,q)$ and $L(p, q^{\ast})$ are Homeomorphic to each other. As expected, the partition function \eqref{det2} is symmetric under $ q \leftrightarrow q^{\ast}$. Moreover, for a given $p$, set of allowed $q$'s is same as the set of $q^{\ast}$'s. The $q$ sum appearing in \eqref{sp3_tot} actually may result in a maximum of double counting.}
\bea \label{sp3_tot}
Z_{\mathrm{tot}} = C_3 \sum_{p=1}^{ \infty}\frac{ e^{2 \pi k /p}}{k^3 \, p^3} \sum _{ \substack{q( \mathrm{mod}\, p)\\ (q,p)=1}} \mbox{ product of sine functions}(q, q^{\ast},p)
\eea
Clearly, the second sum is bounded between $ \pm \phi (p)$, as can be seen from \eqref{totient}. Hence
\bea \label{compare}
|k^3\, Z_{\mathrm{tot}}/C_3| \leq   \sum_{p=1}^{ \infty} \frac{e^{2 \pi k /p}}{p^3} \phi(p)  \leq  \sum_{p=1}^{ \infty} \frac{e^{2 \pi k /p}}{p^2} = \left( \sum_{p=1}^{ p_0} + \sum_{p=p_0+1}^{\infty}\right)\frac{e^{2 \pi k /p}}{p^2}
\eea
where we have split the sum into one up to $p_0$, greatest integer less than the irrational $2 \pi k$, and rest. The splitting is done purely for a matter of convenience and exemplification. In the third step of \eqref{compare}, we have used the fact that the totient function $ \phi(p)$, which counts the natural numbers less than and co-prime to $p$, must be bounded above by $p$. Equality is achieved when $p$ is prime. In the last step, the first finite sum $ \displaystyle{\sum_{p=1}^{ p_0}} \frac{e^{2 \pi k /p}}{p^2}=: M_k$ is definitely finite for whatever large but finite $k$. Hence
\bea
|k^3\, Z_{\mathrm{tot}}/C_3| &< & M_k + e \sum_{p=p_0+1}^{\infty} \frac{1}{p^2} \non \\
						& < & M_k + e \,\zeta(2)
\eea
Hereby we have proved finiteness of the 1-loop partition function, summed over all Lens spaces.

\subsection{Spins$>4$}
Let us now compare the forms of the 1-loop partition functions coming from \eqref{ls1}, \eqref{det2} and \eqref{det4}. These forms are clearly suggestive and allows us to conjecture the structure of the values of the spin-s determinants:
\bea \label{dets}
Z^{[\mathrm{1-loop}]}_{(s)}= \frac{C^{\prime}}{p^2} \, \prod_{r=1}^{s-1}\prod_{\pm} \sin \left( r \pi  \frac{q \pm 1}{p} \right)\sin \left( r \pi  \frac{q^{ \ast } \pm 1}{p} \right)
\eea
The correction by zero modes can be done by a straightforward combinatoric computation, which shows that dimension of traceless Killing tensors (which are assumed to form the kernel of the higher spin ghost operators as well) of all ranks $>2$ all form a 2-dimensional vector space, following the analysis of \cite{Giombi:2013yva}. From this we infer that the zero mode correction for all spins $\sim p/k$, to be multiplied with \eqref{dets}. Therefore, if a truncation of the higher spin tower is made at spin $N$, then the full partition function upto 1-loop quantum correction takes the following form for a generic Lens space:
\bea \label{full}
Z \sim \frac{e^{2\pi k/p}}{(k\,p)^{N-1}}\prod_{\pm} \prod_{s=2}^{N-1} \prod_{r=1}^{s-1} \sin \left( r \pi  \frac{q \pm 1}{p} \right)\sin \left( r \pi  \frac{q^{ \ast } \pm 1}{p} \right)  
\eea
 
\section{Conclusion and outlook}
The principal result of this paper is that spin-4 is the minimum height of the higher spin tower that one has to couple with pure spin-2 gravity in order to get finite 1-loop partition function and hence the normalizable Hartle-Hawking wave-function. We exhibited the calculation in metric and spin-3 Fronsdal variables in most detail. Some of the key points were deriving the required determinant in gauge invariant manner and its evaluation on Lens spaces. Using the known results in harmonic analysis (encoded in the heat Kernel), the determinant was evaluated by a lengthy regularization process. A careful analysis of the zero modes was the next non-trivial step involved. Finally, while summing over all classical saddles we found that the partition is not yet clean from the $\zeta(1)$ divergence encountered in the tree level and 1-loop spin-2 theory. 

As guided by the back-of-the-envelope estimate of the convergence analysis \eqref{formpart} we then ventured the spin-4 sector, which finally paid off. The derivation of the spin-4 1-loop determinant essentially follows the spin-3 one at each step but with increasingly longer equations. Therefore not wishing to clutter the paper with those unnecessary detailed steps, we displayed the final answer for it, even including the analysis of zero modes. An easy convergence test was sufficient to show that a tower of spin-4 is safe for our case. Further we conjecture a form of the partition function for spins greater than 4 as well.

It is tempting to try understand the observed phenomenon of convergence of partition function for higher spins from physical perspectives. As discussed earlier, in 3d it is consistent to have an finite height of higher spin tower coupled to spin-2 gravity, in contrast to 4 dimensions. Even if one understands that interactions like higher spins are key to finiteness of the partition function, it is still baffling to see that spin 4 is the chosen one, not spin 3. Although the Chern Simons answer (appendix C) does not exactly match with the metric-Fronsdal one, it still supports the issue of convergence for spin-4 field and higher.

Deeper understanding of this is one major issue that can and should be ventured in future studies. One of the ways of attacking the problem is try to see the holographic scenario. This may be done by using Maldacena's novel analytic continuation so as to import the results from corresponding AdS/CFT framework \cite{Gaberdiel:2010ar} to Sitter in the spirit of spin-2 case as in \cite{Castro:2012gc}. In that case the wave function depends on the future boundary data of the cosmological space-time. 

Another way to approach is by observing that inclusion of topological parity violating terms \cite{Basu:2011vs} or propagating massive interactions in the theory makes it sensible in terms of convergence properties. In this direction, one may wish to couple higher spins in presence of massive propagating modes and do the computation analogous to the one presented in the main body of this article. On the other hand, the analogous holographic computation can be done by starting with the direct result in AdS \cite{Bagchi:2011td}.

One of the most striking and recently discovered dualities in AdS/CFT correspondence \cite{Gaberdiel:2010pz} is for a (3 dimensional) bulk higher spin (infinite number of higher spins) theory containing massive scalar. This duality has been used recently in understanding the massless modes in tensionless string theory and thus making the expected correspondence between tensionless string theory and a theory of higher spins more concrete. This may be a curious play-ground for the de Sitter analysis. It will be a challenging exercise that, if successful will clarify two issues. The first one is the validity of our analysis or its generalization to infinite higher spins, possibly coupled to new scalars. The second would be to see how far the Maldacena analytic continuation holds. 

Another very important open issue, which is being pursued by the present author is a clear realization of the gauge theory aspect of the analysis, starting from a Chern Simons theory. CS theory gives us a result which manifestly shows the removal of divergence for higher spins with spin 4 or greater. This is a qualitative aspect of the analysis. However more precise questions on the descent of the 1-loop semi-classical result from the non-perturbative one are yet to answered and are being pursued at present.  
\section*{Acknowledgements}
The author would like to thank Arjun Bagchi, Daniel Grumiller and Dmitri Vassilevich for extensive discussions and insightful comments on the manuscript. Discussions with Samir K Paul, Chethan Krishnan, Shiraz Minwalla, Gautam Mandal, Rajesh Gopakumar, Alejandra Castro, Stephane Detournay, Shankhadeep Chakrabortty, Sunil Mukhi, Shailesh Lal, Arunabha Saha, Stefan Fredenhagen and Evgeny Skvortsov are gratefully acknowledged. The author thanks VUT-Vienna, ULB-Brussels and AEI-Potsdam for hospitality and the HoloGrav network of European Science Foundation, Max Planck Group-DST(Govt of India) mobility grant for financial support during the course of this work.
\appendix
\section{Calculation of the Jacobians}
Let us start with the fully decomposed form of the spin-3 field \eqref{split}, \eqref{gaugesplit}, \eqref{splitxi}:
\bea \label{fulldecomp}
\phi _{\mu_1 \mu_2 \mu_3} &=& \phi^{\mathrm{TT}} _{\mu_1 \mu_2 \mu_3}+ g_{(\mu_1 \mu_2}\psi^{\mathrm{T}} _{\mu_3)} + g_{(\mu_1 \mu_2}\nabla _{\mu_3)} \chi+ \nabla_{(\mu_1} \xi^{\mathrm{TT}} _{\mu_2\mu_3)}+ \nabla_{(\mu_1} \nabla_{\mu_2} \sigma^{\mathrm{T}}_{\mu_3)} \non \\
&+& 2(\nabla_{(\mu_1} \nabla_{\mu_2} \nabla_{\mu_3)} - \frac{1}{3}g_{(\mu_1 \mu_2}\nabla _{\mu_3)} \nabla^2 ) \omega 
\eea
The required Jacobian would be calculated through \eqref{defJ}. In the following we would be using the shorthand $\langle A, B \rangle$ for 
$$ \int \sqrt{g} \,d^3 x \, A^{\ast} B_{\ast}$$
where $\ast$ denotes arbitrary index structure, which should be same for both $A$ and $B$ making the product invariant. A very long direct calculation then gives:
\bea \label{inprod}
\langle \phi, \phi \rangle &=&  \langle \phi^{\mathrm{TT}}, \phi^{\mathrm{TT}} \rangle + 15 \langle\psi^{\mathrm{T}},\psi^{\mathrm{T}}\rangle - 15 \langle \chi,\Delta\chi\rangle -3 \langle \xi^{\mathrm{TT}},( \Delta + 6)\xi^{\mathrm{TT}}\rangle \non \\
&+& 12 \langle \psi^{\mathrm{T}}, (\Delta + 2) \sigma ^{\mathrm{T}}\rangle + 48 \langle \sigma ^{\mathrm{T}}, (2\Delta + 3) \sigma ^{\mathrm{T}}\rangle +12 \langle \Delta \sigma ^{\mathrm{T}}, \Delta \sigma ^{\mathrm{T}}\rangle \non \\
&-& 16 \langle \chi , \Delta^2 \omega \rangle - 48\langle \chi , \Delta \omega \rangle - \frac{20}{3} \langle \Delta \omega, \Delta ^2 \omega \rangle +8\langle \Delta \omega, 3 \Delta^2 \omega + 4 \Delta \omega \rangle \non \\ &-& 36 \langle \Delta \omega, \Delta^2 \omega \rangle - 24 \langle \Delta \omega , (9 \Delta +16) \omega\rangle 
\eea

We notice the presence of the disturbing `off-diagonal' terms $\langle \psi^{\mathrm{T}}, (\Delta + 2) \sigma ^{\mathrm{T}}\rangle, ~ \langle \chi, \Delta \omega\rangle$ and $\langle \chi, \Delta^2 \omega\rangle$ in the above inner-product. However, it is easy to figure out that the following linear transformations:
\bea \label{trans}
&& \psi^{\mathrm{T}}_{\mu} \rightarrow \tilde{\psi}^{\mathrm{T}}_{\mu} = \psi^{\mathrm{T}}_{\mu} + \frac{2}{5}( \Delta +2) \sigma ^{\mathrm{T}}_{\mu} , ~~\sigma^{\mathrm{T}}_{\mu} \rightarrow \tilde{\sigma}^{\mathrm{T}}_{\mu} = \sigma^{\mathrm{T}}_{\mu}  \non \\
&& \chi \rightarrow \tilde{\chi} = \chi + \frac{8}{15} ( \Delta +3) \omega , ~~ 
\omega \rightarrow \tilde{\omega} = \omega\eea
remove them. In these new variables, \eqref{inprod} becomes:
\bea \label{chinprod}
\langle \phi, \phi \rangle &=&  \langle \phi^{\mathrm{TT}}, \phi^{\mathrm{TT}} \rangle + 15 \langle\tilde{\psi}^{\mathrm{T}},\tilde{\psi}^{\mathrm{T}}\rangle - 15 \langle \tilde{\chi},\Delta \tilde{\chi}\rangle -3 \langle \xi^{\mathrm{TT}},( \Delta + 6)\xi^{\mathrm{TT}}\rangle \non \\
&+& \frac{48}{5} \langle \tilde{\sigma} ^{\mathrm{T}}, (\Delta + 2)(\Delta + 7) \tilde{\sigma}^{\mathrm{T}}\rangle -\frac{72}{5} \langle \tilde{\omega}, \Delta (\Delta+3) (\Delta +8)\tilde{\omega} \rangle
\eea
The Jacobian factor coming from the linear transformation \eqref{trans} is again trivial, ie
$$  [D \psi^{\mathrm{T}}][D\chi] [D \sigma^{\mathrm{T}}][D \omega] =[D \tilde{\psi}^{\mathrm{T}}][D\tilde{\chi}] [D \tilde{\sigma}^{\mathrm{T}}][D \tilde{\omega}] .$$
Hence, from the definition of the Jacobian \eqref{defJ} and `diagonalized' form of the inner product \eqref{chinprod} we find that:
\bea\label{jacob}
J &=& [\mathrm{det}(-\Delta -6)^{\mathrm{TT}}_{(2)}\, \mathrm{det}\{(-\Delta- 2)_{(1)}^{\mathrm{T}}(-\Delta- 7)^{\mathrm{T}}_{(1)}\} \non \\
&\times & \mathrm{det}\{ (-\Delta)^2_{(0)} (-\Delta -3)_{(0)} (-\Delta -8)_{(0)}\}]^{1/2}
\eea
Finally, in order to match the gauge group volume cancellation, as appearing in \eqref{part} we will have to evaluate the Jacobian for the transformation: $ \xi \rightarrow ( \xi^{\mathrm{TT}}, \sigma^{ \mathrm{T}}, \omega)$ as defined in \eqref{jac2}. As per our usual convention, start by noting that:
$$\int [D\xi] e^{-\langle \xi, \xi \rangle}=1 =\tilde{J} \int [D\xi^{\mathrm{TT}}] [D\sigma^{ \mathrm{T}}] [D\omega] e^{-\langle \xi, \xi \rangle}.$$
Now we can expand the inner product as:
$$\langle \xi, \xi \rangle = \langle \xi^{\mathrm{TT}}, \xi^{\mathrm{TT}} \rangle + 2 \langle \sigma^{ \mathrm{T}}, (-\Delta -2)\sigma^{ \mathrm{T}}\rangle +\frac{8}{3} \langle \omega, (-\Delta)(-\Delta -3)\omega \rangle $$
Hence \bea \label{jac2eval}
\tilde{J}^{-1} &=&\int [D\xi^{\mathrm{TT}}] [D\sigma^{ \mathrm{T}}] [D\omega] e^{-\langle \xi^{\mathrm{TT}}, \xi^{\mathrm{TT}} \rangle - 2 \langle \sigma^{ \mathrm{T}}, (-\Delta -2)\sigma^{ \mathrm{T}}\rangle -\frac{8}{3} \langle \omega, (-\Delta)(-\Delta -3)\omega \rangle} \non \\
&=& [\mathrm{det} (-\Delta -2)^{\mathrm{T}}_{(1)} \mathrm{det}\{ (-\Delta)_{(0)}(-\Delta -3)_{(0)}\}]^{-1/2}
\eea
It follows directly from \eqref{jacob} and \eqref{jac2eval} that:
\bea \label{ratio}
\frac{J}{\tilde{J}} = [\mathrm{det}(-\Delta -6)^{\mathrm{TT}}_{(2)}\, \mathrm{det}(-\Delta- 7)^{\mathrm{T}}_{(1)} \mathrm{det}\{ (-\Delta)_{(0)} (-\Delta -8)_{(0)}\}]^{1/2}
\eea
\section{Resummation of the `oscillatory' zeta function} \label{app2}
For the purpose of exemplification, let's consider from the expression \eqref{semifexprn}, the term 
\bea \label{resum}
&&\sum_{m \in \mathbb{Z}_p} \cos( m \tau) \tilde{\zeta}_{0}(m \theta_1 ; z) \non \\
&&= \frac{1}{2} \sum _{n=1}^{ \infty} \frac{1}{n^z} \sum_{m \in \mathbb{Z}_p}\left[ \cos\left(\frac{2 \pi m}{p} \left(n q +q-1 \right)\right)+\cos\left(\frac{2 \pi m}{p} \left(n q -q+1 \right)\right)\right] \non\\
&& = \frac{1}{2} \sum _{n=1}^{ \infty} \frac{1}{n^z}\sum_{m \in \mathbb{Z}_p}\left[ \cos\left(\frac{2 \pi m q}{p} \left(n +1-\q \right)\right)+\cos\left(\frac{2 \pi m q}{p} \left(n  -1+\q \right)\right)\right]
\eea
Here we have defined the integer $ \q$ (mod $p$) such that $q \q =1 \, (\mathrm{mod}\,p)$. The $m$ sum for the first term of the last line of \eqref{resum} above would give $\displaystyle{\sum_{k \in \mathbb{Z}}} \delta _{n+ 1- \q, pk}$. This Kronecker delta would therefore result in a resummation of the infinite $n$ sum. This makes the resummed \eqref{resum} more tractable:
\bea \label{Hurwitz}
&&\frac{p^{1-z}}{2} \left[ \sum_{k=0}^{ \infty} \frac{1}{\left( k + \frac{\q -1}{p}\right)^z}+ \frac{1}{\left( k - \frac{\q -1}{p}\right)^z} - \left(- \frac{p}{ \q -1} \right)^z \right] \non \\
&=& \frac{p^{1-z}}{2} \left[ \zeta (z,\frac{\q -1}{p} ) + \zeta (z,-\frac{\q -1}{p} ) - \left(- \frac{p}{ \q -1} \right)^z\right]
\eea
Where we have used the definition of more familiar Hurwitz zeta functions:
$$ \zeta (z,a) = \sum_{n=0}^{\infty} \frac{1}{(n+a)^z}$$
for Re($z)>1$ and Re($a)>-1$ and $ \neq 0$. The properties that are of importance in our calculation is that:
$$ \zeta(0,a) = \frac{1}{2}-a, ~~\zeta^{\prime}(0,a)+\zeta^{\prime}(0,-a) = - \ln (- 2a\,\sin ( \pi a)) $$
Now as it appears in \eqref{semifexprn}, we need to act $ - \frac{2}{p} \frac{d}{dz}$ on \eqref{Hurwitz} and evaluate at $z=0$. The result is:
\bea
 \ln \left(2\sin \left( \pi \frac{\q-1}{p}\right) \right)
\eea
\section{Results in Chern Simons formulation}
In this Appendix we describe the partition function up to spin-3 dynamics, described in terms of $SU(3)\times SU(3)$ Chern Simons theory. To set the stage, we will briefly review the technique of non-perturbative quantization of Chern Simons theory on Lens spaces.
\subsection{Background}
Euclidean dS${}_3$ pure gravity is equivalent to $SU(2) \times SU(2)$ Chern Simons (CS) theory at level $k$:
\bea \label{csaction}
S[A^+ ,A^-] &=& \frac{k}{4 \pi} \mathrm{tr}\int _M ( A^+ \wedge d A^+ + \frac{2}{3} A^+ \wedge A^+ \wedge A^+)\nonumber \\&-&\frac{k}{4 \pi} \mathrm{tr}\int _M ( A^- \wedge d A^- + \frac{2}{3} A^- \wedge A^- \wedge A^-)
\eea
where $A^{\pm} = \omega \pm e$ are $ \mathfrak{su}(2)$ valued connections. $k=\frac{1}{4G}$ in our system of units that sets cosmological constant to unit. Since $M$, the base manifold is closed (static patch of Euclidean dS${}_3 \sim S^3$ or $S^3 / \Gamma$), ie $\partial M = \emptyset $, we need not bother about boundary terms for variational principle to hold. The action is also finite for all classical saddles. By equation of motion, all solutions are flat connection. 

Since the background manifold $M\equiv L(p,q)$ is closed, it may be natural to look for canonical quantization on physical phase space of a CS theory. However, a first attempt at straightforward geometric quantization \cite{Axelrod:1989xt, Elitzur:1989nr} of the theory poses with subtle challenges. As there are no local degrees of freedom and the solution space consists of only flat connections, the essential data about Lens space that suffices this purpose is its topology. The topological structure of $L(p,q) = S^3 / {\mathbb Z}_p$ has been explained in the Introduction section in some detail. The phase space of physically inequivalent solutions of this theory consists of homomorphism maps from its first homotopy group to the gauge group, modulo gauge covariance: $\left( \hom : \pi _1 \left(L(p,q)\right) \rightarrow SU(2) \right)/\sim$, (in other words, moduli space of flat $\mathfrak{su}(2)$ valued connections modulo gauge transformations). Here $\sim$ denotes group conjugation. In physical terms, this is a coordinatization of the phase space by non-trivial Wilson lines. The fundamental group of the Lens space $L(p,q)$, is ${\mathbb Z}_p$. This is Abelian and consists of words freely generated by a unique letter, say $ \alpha $. Therefore this group has the elements $\{ \alpha ^n | n=0,\dots ,p-1\}$. Let $h$ denote the maps from ${\mathbb Z}_p$ to the gauge group $SU(2)$. For $h$ to be homomorphism, it should obey the following rule:
 \bea \label{conjugation}h[ \alpha ^p] = \left(h[ \alpha] \right)^p = \mathbf{1}.\eea
$\mathbf{1}$ is the identity element of $SU(2)$. For ease of exemplification, let's use the two dimensional $j=1/2$ unitary representation of $SU(2)$. In this representation, using the freedom of conjugation of group elements, we can choose an unique solution of \eqref{conjugation} (as there is only one conjugacy class in this group) as: 
$$ h[ \alpha] = \exp(2\pi i \sigma _3 /p).$$
Therefore the phase space or moduli space is a collection of $p$ distinct points. It is not even a manifold, leave apart symplectic. In other words, these points correspond to independent holonomies of $p$ disjoint non contractible loops, which can have only discrete values.

Note that while constructing the space of gauge inequivalent solutions through Wilson lines, the detailed structure of the flat connections were glossed over. For example the configuration corresponding to identity element of the group above can be trivially given by the empty solution $A^{\pm} =0 $ or its gauge equivalent class. While thought in terms of gravity, a pair of this kind of solutions from each of the $\pm$ sectors of the theory \eqref{csaction} would give rise to $e=0=\omega$. Of course these dreibeins are not invertible. This is a unique situation in first order formulation of gravity and does not give rise to any meaningful gravitational configuration, as explained in detail in \cite{Witten:2007kt}. But if one wishes to perform non-perturbative quantization of the first order theory, there is no apparent reason for excluding them beforehand as well.

Now as we have seen that a canonical phase space of the theory does not exist for Lens space topologies, there is no scope of canonical quantization. However, the innovative techniques of gluing and pasting developed in \cite{Witten:1988hf} gives one handle \cite{Jeffrey:1992tk} to calculate partition function of the theory non-perturbatively and hence complete the quantization. As explained in the introduction, we note that $L(p,q)$ can be constructed by `properly' gluing two solid tori at their boundary. By `properly', we mean the action of 
\[ U_{p,q}=\left( \begin{array}{cc}
q & b  \\
p & d  \end{array} \right) \in PSL(2, \mbz)\]
on the boundary of one of the tori before gluing them together. One then goes on canonically quantizing the CS theory on the torus boundaries with respect to the symplectic structure induced on its finite dimensional physical phase space. This gives rise to $k+1$ dimensional Hilbert space of quantum states, where $k$ is the integer CS level. This Hilbert space can be conveniently characterized by Weyl-reflection invariant $SU(2)_k$ Weyl-Kac characters:
\bea \label{theta_fn}
\psi _{ j , k} (z, \tau) = \frac{\vartheta _{j , k+2} (z , \tau) -\vartheta _{-j , k+2} (z , \tau)}{\vartheta _{1 , 2} (z , \tau) -\vartheta _{-1 , 2} (z , \tau)} ~~~~~~~~ j = 0 , \dots k
\eea
An easily accessible derivation of this can be found in \cite{Basu:2011vs}. $z$ is the holomorphic coordinate and $\tau $ is the modular parameter on the moduli space of the flat connections. Now, the action of $PSL(2, \mathbb{Z})$, at least the representation of its generators $S$ and $T$ on this vector space is well known. However, we must now remember the fact that we must quantize on the solid torus. The state space \eqref{theta_fn} applies to 3-manifolds of the form circle bundles which can be locally trivialized as $T^2 \times S^1$. Going to the solid torus, one projects out all the states from the above Hilbert space to a single state. If we consider a Wilson line carrying representation $j \in \mathbb{N}$ of $SU(2)$ inserted along the non-contractible direction in the solid torus, the only state corresponding to that would be $\psi _{ j , k} (z, \tau)$ above \cite{Labastida:1989xp}.

Using this information and picking up the solid torus with no Wilson line inserted (or inserted in trivial representation) as described by wave-function $\psi _{ 0 , k}(z,\tau)$ , one calculates the following matrix element:
$$Z= \langle \psi _{ 0 , k} | U_{p,q} | \psi _{ 0 , k} \rangle $$ 
which is known as Reshetikhin-Turaev invariant \cite{Hansen:2001kp} in the mathematical literature. Its physical interpretation is the non-perturbative partition function. In a special choice of `framing' the above matrix element is evaluated \cite{Jeffrey:1992tk} to be
\bea \label{Z1}
Z &=& -\frac{i}{\sqrt{2(k+2)p}}\exp \left( 6 \pi i S(p,q) /(k+2)\right) \nonumber \\&\times & \sum _{\pm} \sum _{n=1}^{p} \exp \left( \frac{2\pi iq (k+2) n^2}{p} + \frac{2 \pi i n (q \pm 1)}{p} \pm \frac{\pi i}{(k+2)p}\right)
\eea
where,
$$S(p,q) = \sum _{l=1} ^{p-1} \dfrac{l}{p} \left( \dfrac{l q}{p }- \left\lfloor \dfrac{lq}{p}\right\rfloor - \dfrac{1}{2}\right)$$ is the Dedekind sum and $\lfloor x \rfloor$ is the floor function of $x$ (highest integer less than $x$). From the expression pf \eqref{Z1}, it becomes clear that the theory is 2-loop exact, as the terms in the exponential stops at $\sim (k+2)^{-1}$ in inverse power series of $k+2$. Interestingly one also notes the term $\sim \frac{1}{2}\log(k+2)$ in the same series, which manifests as the $\frac{1}{\sqrt{(k+2)}}$ factor in \eqref{Z1}. This is the non-perturbative instanton effect. This was introduced in the perturbative metric formalism through the zero mode contribution, cf. \eqref{ls1}, \eqref{zm}.

The large-$k$ limit of \eqref{Z1} is the semi-classical one and is equivalent to what is understood as the 1-loop partition function. The $SU(2)$ large level-$k$ CS partition function takes the following form on Lens space $L(p,q)$:
\bea \label{CS}
Z = i \,\sqrt{\frac{2}{(k+2)p}} \sum_{n=1}^{p} e^{2 \pi i (k+2) n^2 \frac{q^{\ast}}{p}} \sin\left(2 \pi n \frac{q^{\ast}}{p} \right)\, \sin\left(2 \pi \frac{n}{p} \right).
\eea
One might notice the quantum level shift $k \rightarrow k+2 $ by dual Coxeter number that is omnipresent in the theory of Lie-group symmetric CFTs, Chern Simons theory Hilbert space \eqref{theta_fn} analysis is also present in the formula \eqref{CS}. In the large $k$ limit this is not of big significance but will be important in the large $N$ (of $SU(N)$) limit. Note that each of the terms of the $n$-sum in the above formula actually denotes a classical CS saddle corresponding to $p$ disjoint gauge invariant moduli class of flat connections (solution of the CS e.o.m: $F =0$) on $L(p,q)$.

In an attempt to match the CS result to the gravity one we should follow the steps outlined below, (assuming that we are in the large $|k|$ regime).
\begin{enumerate}
\item
Take product of two such partition functions \eqref{CS}, each from one $SU(2)$ sector having equal and opposite levels $k$ and $-k$. However we remind the reader of a caveat here. The quantization of $SU(2)$ CS, which leads to the partition function \eqref{CS} on Lens spaces, severely restricts $k$ to non-negative integers \footnote{That $k$ must be an integer can be seen even before quantization, ie the pre-quantization condition on the integrality of the sympletic structure, on the phase space, which is finite dimensional, compact and closed}, since the Hilbert space of physical states is $k+1$ dimensional, as seen from \eqref{theta_fn}. In this sense validity of the above formula \eqref{CS} for both $\pm k$ can only be taken in the analytically continued sense, believing that such a continuation exists.
\item
Neglect the level shift $\pm k \rightarrow \pm k +2$. This is well justified since we are in the large $|k|$ regime. We would like to remark here that, if we consider a very `high' tower of spins, described by large $N$ (of $SU(N)$), the shift $\pm k \rightarrow \pm k +N$ may not be negligible. 
\item
The sum over $n$ in \eqref{CS} is actually a sum over the discrete moduli-space of flat $SU(2)$ bundles on $L(p,q)$. In physical terms these are gauge invariant saddles of the theory. Out of the $p$ saddles appearing in the sum, a particular saddle has to be chosen from each of the two $SU(2)$ sides,  \footnote{Note that again we risk running into a meaningless regime, since $\frac{q \pm 1}{2}$ is not always an integer. This formula also holds in the sense of liberty coming with the power of analytic continuation} namely: $n_{\pm} = \frac{q \pm 1}{2}$. The flat connections corresponding to these two moduli numbers produce dS${}_3$ geometry with no conical defect through identifying the metric as $g = \frac{1}{4}\mathrm{tr}[(A^+ -A^-)\otimes(A^+ -A^-)]$. 
\item
Finally the analytic continuation of the CS level: $k \rightarrow -ik$ has to be made so as to make the Euclidean CS action becomes equivalent to the the Einstein Hilbert action. 
\end{enumerate}
With these considerations, one can reproduce the spin $N=2$ special case of \eqref{ls1}.

\subsection{Spin-3 (non)-perturbative CS result}
AdS${}_3$ pure gravity (in Lorentzian signature) is described by $SL(2, \mathbb{R}) \times SL(2, \mathbb{R})$ CS and all spins upto $N$ are included in $SL(N, \mathbb{R}) \times SL(N, \mathbb{R})$ description. We also know that both the groups $SU(N)$ and $SL(N, \mathbb{R})$ are built upon the same Lie-algebra $A_{N-1}$ with identical root-system. Hence the natural choice of extending $SU(2) \times SU(2)$ to incorporate higher spin algebras for de Sitter is to choose $SU(N) \times SU(N)$ CS and identify the principally embedded $\mathfrak{su}(2)$ :$$ \mathfrak{su}(N)=\mathfrak{su}(2) \oplus \mathbf{5} \oplus \mathbf{7}\cdots $$ as describing the pure spin-2 gravity. With this motivation, we consider $ SU(N) \times SU(N)$ CS at level $\pm k$ for describing an interaction of all spins. We will be using the notations of \cite{Campoleoni:2010zq}. For specializing to spin-3, we take $N=3$. 

The CS quantization scheme on $L(p,q)$ essentially follows the same routine as in the $SU(2)$ case. The Hilber space of $SU(3)$ CS on a Seifert fibered \cite{Lawrence:1999rp} space $T^2 \times S^1$ can be characterized by a convenient basis for of the $SU(3)_k$ Weyl-Kac characters $\psi_{\lambda,k}$. $\lambda$ are all weights of $A_2$ root-system satisfying $$k \geq (\lambda, \alpha) \geq 0 $$ 
for the three positive roots \cite{Axelrod:1989xt}. This inequality can be solved to give $(k+1)(k+2)/2$ weights $\lambda$, which therefore becomes the dimension of the Hilbert space. 

As per the prescription in the $SU(2)$ case, we consider the state corresponding to $\lambda=0$ defining the solid torus. The non-perturbative partition function for $SU(3)$ CS on $L(p,q)$ at level $k$ (shifted to $k^{\prime} =k+3$) is thus given by the element $\langle \psi_{\lambda=0,k} |U_{p,q}|\psi_{\lambda=0,k}\rangle$:
\bea \label{CS3np}
Z &=& \frac{1}{\sqrt{3}\, pk^{\prime}}e^{\frac{2 \pi i}{k^{\prime}}S(q,p)} \sum_{n_1 , n_2 \in \mathbb{Z}_p} \exp \left(\frac{2 \pi i q}{p} \left(k^{\prime} \left( n_1^2+n_2^2 - n_1 n_2\right)+ n_1 +n_2 \right)\right) \non \\
& \times & \left[ \sin \left(\frac{\pi}{p}\left( n_1+n_2 + \frac{2}{k^{\prime}}\right)\right)+\sin \left(\frac{\pi}{p}\left( n_1-2 n_2 - \frac{1}{k^{\prime}}\right)\right)-\sin \left(\frac{\pi}{p}\left( 2 n_1-n_2 + \frac{1}{k^{\prime}}\right)\right)\right].
\eea
But at large $k$ its $S(q,p)$ dependence goes away \cite{Hansen:2003ky} as in the $SU(2)$ case \eqref{CS} and the shift $k \rightarrow k+3$ can be neglected:
\bea \label{CS3p}
Z=\frac{8}{\sqrt{3}\,k p} \sum_{ \substack{\{n_1,n_2\} \in \\ \mathbb{Z}_p \times \mathbb{Z}_p}} && \exp \left(\frac{2 \pi i k q^{\ast}}{p} \left( n_1^2+n_2^2 - n_1 n_2\right) \right) \sin \left(\frac{2 \pi q^{\ast}}{p} \left(n_1 +n_2 \right)\right) \non \\
&\times & \sin\left(\frac{\pi}{p}\left( n_1+n_2\right) \right)\sin\left(\frac{\pi}{p}\left( 2n_1-n_2\right) \right)\sin\left(\frac{\pi}{p}\left( n_1-2n_2\right) \right)
\eea
It is interesting to note that the summand above picks up non-zero value for terms for which none of the combinations $n_1 +n_2,\, 2n_1 -n_2,\, n_1 -2 n_2$ evaluate to be $0$ or $p$.

For constructing the 1-loop partition function of gravity coupled with spin-3 fields from the information \eqref{CS3p}, we must perform the steps enlisted at the end of the last subsection. Steps 1 to 3 can be trivially implemented for $SU(3)$ starting from \eqref{CS3p}. Also note that the step 1 is free from the subtlety of the $SU(2)$ case. There are no restrictions on positivity of $k$ in this case, as the dimension of the Hilbert space of the $ T^2 \times S^1$ theory, $(k+1)(k+2)/2 $ is positive integer for all integers $k \neq -1, -2$.

Following the first 2 steps outlined in the last section, we do not land up to a very enlightening result. We now turn to implementing step 3. For that, we start by noticing that the sum in \eqref{CS3p} is over a the discrete moduli space of $p^2$ points $(n_1,n_2) \in \mathbb{Z}_p \times \mathbb{Z}_p$ corresponding to the gauge independent flat connections on $SU(3)$ bundles over $L(p,q)$. Each of these integer pair $(n_1, n_2)$ are points on a two dimensional sub-lattice \cite{Hansen:2003ky} of the coroot lattice of $A_2$ system. In physical terms, these correspond to gauge invariant saddles of the theory.

Implementation of step 3 will require identifying a particular saddle from the above mentioned $p^2$ points on the coroot lattice. This should be done so that the final expression should match the gravity side result. Our intuitions from the gravitational aspect of the theory should guide us choosing the a saddle which should be preferred over the other ones. To be more precise, this should correspond to (gauge equivalent class of) a pair of flat connections whose spin-2 part is equivalent to smooth dS${}_3$ metric and spin-3 part switched off to zero. By spin-2 part of the $\mathfrak{su}(3)$ connection we mean the projection of it on the $\mathfrak{su}(2)$ subalgebra, having principal embedding in $\mathfrak{su}(3)$. In this point of view, spin-3 fluctuations start contributing only at the 1-loop quantum level.

This choice amounts to setting either $n_1$ or $n_2$ zero (since the formula \eqref{CS3p} is symmetric under $n_1 \leftrightarrow n_2$). Let's take $n_2 = 0 $ for our case. We have to make this choice for both the C-S sectors of the action \eqref{CS}. For $n_1$ we see that the choice $q+1$ and $q-1$ respectively for the $+$ and $-$ sector of the theory gives us the pure dS${}_3$ background with spin-3 field turned off.

After doing this, we then perform the last step of analytically continuing $k \rightarrow -ik$ to get:
\bea \label{CSfinal}
Z &=& \frac{2^{10}}{3\,k^2 p^2} e^{2 \pi k/p} \sin\left( 2 \pi  \frac{q^{\ast}+1}{p} \right)\sin\left(2 \pi  \frac{q^{\ast}-1}{p} \right)\sin\left(2 \pi  \frac{q+1}{p} \right)\sin\left(2 \pi  \frac{q-1}{p} \right)  \non \\ 
& \times & \sin ^2\left( \pi  \frac{q+1}{p} \right)\sin^2\left( \pi  \frac{q-1}{p} \right)
\eea
This is the final expression that we would like to compare with the metric one \eqref{sp3_pq}. Leaving apart the unimportant numerical factors, the expressions do not match. The above CS result misses the factor $\sin ^2\left( \pi  \frac{q^{\ast}+1}{p} \right)\sin^2\left( \pi  \frac{q^{\ast}-1}{p} \right)$, which prevents the match. However, the $1/p^2$ dependence, which is crucial for the convergence issue, while summing over all Lens spaces, do match. This mismatch is the precise reason that we do not go on to explore the spin-4 case described by $SU(4)\times SU(4)$ CS.

However we must mention here that the partition function for $SU(N)$ on Lens space $L(p,q)$ behaves as $(k\,p)^{-(N-1)/2}$ as far the power law dependence goes, modulo the multiplicative sine functions \cite{Hansen:2003ky}. Product of two copies of $SU(N)$ theory thus results in the desired behaviour of $ \dfrac{1}{(k\,p)^{(N-1)}}$ as in the metric-higher spin formulation \eqref{sp3_pq}, \eqref{spin-4} and \eqref{full}.

\end{document}